\documentclass[secnumarabic, graphics,floatfix, nofootinbib,tightenlines,nobibnotes, aps, prl, 12pt]{revtex4-2}
\usepackage[english]{babel}
\usepackage[utf8]{inputenc}
\usepackage{amsfonts}
\usepackage{multirow}
\usepackage{graphicx}
\usepackage{epstopdf}
\usepackage{hyperref}
\usepackage{latexsym}
\usepackage{amsmath}
\usepackage{amssymb}

\usepackage[utf8]{inputenc}
\hypersetup{
    colorlinks=true,
    linkcolor=red,
    citecolor=blue,
}
\usepackage{color}
\usepackage[T1]{fontenc}
\usepackage{txfonts}
\usepackage{mathrsfs}
\usepackage[center]{subfigure}

\begin{document}
 \setcounter{secnumdepth}{2}
 \newcommand{\bq}{\begin{equation}}
 \newcommand{\eq}{\end{equation}}
 \newcommand{\bqn}{\begin{eqnarray}}
 \newcommand{\eqn}{\end{eqnarray}}
 \newcommand{\nb}{\nonumber}
 \newcommand{\lb}{\label}
 
\title{Qusinormal oscillations and late-time tail of massless scalar perturbations of a magnetized black hole in Rastall gravity}

\author{Cai-Ying Shao$^{1}$}
\author{Yu-Jie Tan$^{1}$}
\author{Cheng-Gang Shao$^{1}$}\email[E-mail: ]{cgshao@hust.edu.cn}
\author{Kai Lin$^{2,3}$}\email[E-mail: ]{lk314159@hotmail.com}
\author{Wei-Liang Qian$^{3,4,5}$}\email[E-mail: ]{wlqian@usp.br}

\affiliation{$^{1}$ MOE Key Laboratory of Fundamental Physical Quantities Measurement, Hubei Key Laboratory of Gravitation and Quantum Physics, PGMF, and School of Physics, Huazhong University of Science and Technology, 430074, Wuhan, Hubei, China}
\affiliation{$^{2}$ Hubei Subsurface Multi-scale Imaging Key Laboratory, Institute of Geophysics and Geomatics, China University of Geosciences, 430074, Wuhan, Hubei, China}
\affiliation{$^{3}$ Escola de Engenharia de Lorena, Universidade de S\~ao Paulo, 12602-810, Lorena, SP, Brazil}
\affiliation{$^{4}$ Faculdade de Engenharia de Guaratinguet\'a, Universidade Estadual Paulista, 12516-410, Guaratinguet\'a, SP, Brazil}
\affiliation{$^{5}$ Center for Gravitation and Cosmology, School of Physical Science and Technology, Yangzhou University, 225002, Yangzhou, Jiangsu, China}

\date{May 15th, 2022}

\begin{abstract}
In this paper, we study the quasinormal mode and late-time tail of charged massless scalar perturbations of a black hole in the generalized Rastall gravity. 
The black hole metric in question is spherically symmetric, accompanied by a power-Maxwell field surrounded by a quintessence fluid.
It is shown that the massless scalar field, when {\it dressed up} with the magnetic field, acquires an effective mass which, in turn, significantly affects the properties of the resultant quasinormal oscillations and late-time tails.
To be specific, the quasinormal frequencies become distorted and might even be unstable for particular spacetime configurations.
Also, the exponent of the usual power-law tail is modified in accordance with the modification in the structure of the branch cut of the retarded Green's function.
In particular, as the effective mass is generated dynamically due to the presence of the magnetic field, one may consider a process through which the field is gradually removed from the spacetime configuration.
In this context, while the quasinormal oscillations converge to the case of massless perturbations, we argue that the properties of resultant late-time tails do not fall back to their massless counterpart.
The relevant features are investigated by numerical and analytic approaches.
\end{abstract}

\maketitle

\newpage
\section{Introduction}

The black hole is one of the most intriguing predictions of Einstein's general relativity.
It is a particular spacetime region typically created by the gravitational collapse of a compact object in its late stage, whose mass exceeds the upper limit of neutron stars~\cite{book-bh11}. 
From both theoretical and experimental perspectives, the subject has received much attention in the past few decades.
As the culmination of continuous efforts, the recent advent of empirical detection regarding the gravitational wave emanated from the black holes has given rise to further incitement.
In particular, the merger signals of a black hole binary captured by the LIGO and Virgo collaboration~\cite{prl-observation-gravitational-waves} furnished the first direct piece of evidence of black holes and has subsequently inaugurated a novel era of gravitational-wave astronomy. 

The quasinormal modes (QNMs) carry the {\it characteristic sound} of a black hole.
In terms of dissipative oscillations, they bear the essential properties of the underlying spacetime metric~\cite{cqg-Nollert-qnm-charac}.
The associated temporal profile constitutes the ringdown phase of the collapse, from which a black hole is formed. 
On the one hand, such signals are mathematically straightforward compared to those emanating from the merger process. 
On the other, the strength of the associated gravitational radiation is speculated to still fall within the scope of the ongoing space-borne gravitational wave detector programs, such as LISA and TianQin~\cite{agr-LISA-01,cqg-tianqin}.
In particular, studies have been performed to evaluate the detector signal-to-noise ratio regarding the feasibility of an eventual detection~\cite{prd-gravit-wave-bh,prd-nonsingula-gravit-pertur}.

On the theoretical side, the QNMs can be investigated using the black hole perturbation theory.
Among the earliest endeavors, Regge and Wheeler laid down the framework of the metric perturbations in spherically symmetric black holes~\cite{prd-stab-01}.
Consequential developments by Zerill~\cite{prl-pertur-equa} and Moncrief~\cite{ap-gravita-perturb} further refined a few pertinent aspects.
Regarding the rotating black holes, Teukolsky first derived the linearized scalar and metric perturbation equations for the Kerr spacetime~\cite{taj-perturbation-rotating, an-math-BH}.
Although the physical system is dissipative and therefore non-Hermitian, the mathematical procedure to obtain the QNMs can be effectively viewed as an eigenvalue problem. 
Subsequently, the emergence of discrete complex frequencies is attributed to the ingoing and outgoing boundary conditions, respectively, applied at the horizon and outer spatial bound (often located at the infinity).
By using the continued fraction method proposed by Leaver~\cite{agr-qnm-continued-fraction-01, agr-qnm-11}, one encounters the discrete complex frequencies reminiscent of the atomic spectrum of hydrogen in quantum mechanics.
The eigenvalue nature of the problem is also transparently seen in terms of the matrix method~\cite{agr-qnm-lq-matrix-01,agr-qnm-lq-matrix-02,agr-qnm-lq-matrix-03,agr-qnm-lq-matrix-04}, where one explicitly solves a matrix secular equation for the quasinormal frequencies.
On a different footing, Leaver proposed Green's function approach, which reformulates the general solution in terms of the spectrum decomposition~\cite{prd-decomp-perturb,cqg-Nollert-qnm-charac}.
From this viewpoint, the main characteristics of the perturbation fields are attributed to the intrinsic singularities of the relevant Green's function.
Specifically, the QNMs correspond to the poles of Green's function, while the branch cuts essentially govern the late-time tails.
Besides the QNMs, the late-time tail is also a topic of pertinent relevance.
On the one hand, the time profile of QNM decays exponentially, while on the other, the late-time tail falls largely following a power-law form. 
Therefore, the last stage of the temporal evolution of the perturbations is dominated mainly by the late-time tail.
As mentioned above, mathematically, the origin of the late-time tail is attributed to the branch cut and cannot be obtained by summing up the QNM poles.
For instance, in the case of massless scalar perturbations, the tail is due to a branch cut on the negative imaginary axis in the frequency domain, which stretches out from the origin.
It has been argued that the presence of the branch cut sensitively depends on the asymptotical structure of the potential~\cite{prd-late-time-stell}.
Also, it is often interpreted in the literature as the backscattering of perturbed wave packets by the spacetime far away from the horizon~\cite{agr-qnm-tail-06,mpla-tail-scalar-global-monopole,mpla-dirac-qnm}.
It was Price who first observed an inverse power-law tail, ${t^{ - (2l + 3)}}$ or ${t^{ - (2l + 2)}}$, for massless scalar perturbations in the Schwarzschild black hole metric~\cite{prd-nonspherical-scalar-gravita}.
Extensive studies for different metrics further indicated that such an inverse power-law form is a rather general feature for massless perturbations in spherical spacetimes which are either asymptotically flat~\cite{agr-qnm-tail-06, agr-qnm-tail-09, agr-qnm-tail-12}.
On the other hand, the temporal profile for massive perturbations might be qualitatively different from their massless counterparts.
For insignificant mass, Hod and Piran discovered that the time-domain profile at the intermediate-late time in Reissner-Nordstr\"om spacetime possesses an oscillatory tail with a decay rate of ${t^{ - (l + 3/2)}}\sin (\mu t)$~\cite{prd-tail-gravi-collap}, where $\mu $ is the mass of the field.
Moreover, Koyama and Tomimatsu showed analytically that the asymptotical late-time tail in the Schwarzschild spacetime reads ${t^{ - 5/6}}\sin (\mu t)$~\cite{prd-tails-rn, Koyama2001}.
For some specific cases, it was recently pointed out that the asymptotical form could also be ${t^{ - 1}}\sin (\mu t)$~\cite{agr-qnm-tail-39} in the limit of vanishing scalar mass.
In all the massive cases, the branch cut can be conveniently chosen to be on the real axis in a finite interval determined by the scalar mass. 
It is also worth noting that such a decay rate does not depend on the angular momentum $l$.
Other than the power-law form, the exponential tails have also been found in asymptotically de Sitter spacetimes for various types of fields~\cite{prd-tails-cosmolog,prd-sch-desitter,prd-de-Sitter}.
It is notable that the Dirac perturbations eventually decay to a non-zero constant~\cite{mpla-tail-dirac,mpla-dirac-qnm}.
These distinctive characteristics have significantly enriched the physical content regarding the late-time waveforms.
More recently, Cardoso {\it et al.} proposed~\cite{prl-gravita-ringdown} that the late-time ringdown may serve to probe the horizon and discriminate between different gravitational systems, inclusively, the exotic compact objects such as gravastar and wormhole.
Following this line, the studies of black hole echoes~\cite{sc-echo-bh}, and particularly, their connections with the QNMs and late-time tail~\cite{prd-mechanism-bh-echoes,prd-recipe-echoes} have also aroused much interest.

The magnetic field is yet another indispensable assembly piece of the Universe.
On the experimental side, employing the rotation measures, information on the magnetic field in the disk of galaxies can be extracted for a large number of pulsars~\cite{cjaas-magnetic}.
In particular, strong magnetic fields, as high as ${\rm{1}}{{\rm{0}}^{\rm{4}}} \sim {\rm{1}}{{\rm{0}}^{\rm{8}}}$ G, have been observed in the vicinity of stellar-mass or supermassive black holes~\cite{piotrovich2010magnetic1}.
Therefore, relevant studies of black hole metrics, and inclusively their perturbations, have to be put in the context of a realistic astrophysical environment.
A stationary axisymmetric black hole solution with a uniform magnetic field placed along the symmetry axis was first derived by Wald~\cite{prd-bh-uniform-magnetic}. 
The solution for the Kerr-Newman black hole was subsequently obtained by Ernst and Wild~\cite{jomp-01}.
Among many important developments, the magnetic field is found to play an essential role in the stability of the dynamic system.
The latter encompass the superradiance instability~\cite{prd-superradiant-instability} and black hole QNMs~\cite{plb-qnm-magn, prd-qnm-magn-bh}.
In particular, the Zeeman effect was recently uncovered in the QNM spectrum of charged scalar perturbations in a magnetized Schwarzschild black hole~\cite{plb-qnm-magn}.
Moreover, instability of the metric can be triggered by tuning the effective scalar mass, which is modified due to the interaction between the charge and magnetic field.

As an alternative theory of general relativity, Rastall gravity was formulated based on the possible breaking of the conservation law of the stress energy-momentum tensor~\cite{prd-gener-einstein}.
The theory is aimed to address several unsettled issues in the general relativity regarding the largest scale when compared against cosmological observation.
In the literature, various intriguing aspects of the theory have been explored, which include black hole physics~\cite{cjp-bh-rastall,epjc-rotati-bh,ijmpd-therm-bh-rastall,rotating-bh-shadow} and cosmology~\cite{jp-rastall-cosm-03,epjc-genera-rastall,epjc-observ-constra-rasta,plb-therm-flat-ras,plb-resolu-cosmo}.
Among others, the theory inherently gives rise to a particle creation mechanism, which naturally supplies an alternative implementation for the dark energy~\cite{epjc-cosmic-dark-energy-rastall}. 
It is meaningful to investigate further the role of the magnetic field in Rastall theory regarding its impact on spacetime stability.

The present work is motivated by the above considerations.
We investigate the QNMs and late-time tails of charged massless scalar perturbations of a magnetized black hole solution derived from generalized Rastall gravity.
First, we generalize the black hole solution in Rastall gravity into a more realistic context by including the magnetic field.
We then analyze how the magnetic field will affect the spacetime stability in the context of Rastall gravity by exploring the quasinormal oscillations.
For the latter, as it turns out, the study of massless scalar perturbations suffices for the purpose.
For stable spacetime configurations, it is understood that massless and massive scalar fields behave distinctively in the late-time evolution.
Therefore, in the second place, it would be rather interesting to investigate what would be the case for Rastall's theory with the presence of a magnetic field.
In particular, we find that the exponent of the power-law tail is modified.
This can be attributed to the fact that the effective mass of the field is ``dressed up'' owing to the interaction between the magnetic field and the charge.

The remainder of the paper is organized as follows.
In the next section, we present the metric of a magnetized black hole investigated in the present study.
The QNMs are explored in Sec.~\ref{section3} by employing the WKB approximation, the matrix method, and the finite difference method. 
In Sec.~\ref{section4}, we proceed to analyze the late-time tail of the massless scalar perturbations and the effect of the magnetic field.
Both Green's function and finite difference method are utilized.
Further discussions and concluding remarks are given in the last section.

\section{Power-Maxwell charged black hole solutions in Rastall gravity } \lb{section2}

According to Rastall's recipe~\cite{prd-gener-einstein}, the breaking of the conservation law of the stress energy-momentum tensor reads
\begin{equation}\label{N1}
{T^v}_{\mu ;v} = \lambda {R_{,\mu }}\ ,
\end{equation}
where $\lambda $ is the Rastall coupling parameter and $R$ is the Ricci scalar.
Obviously, if $\lambda  = 0$, the stress energy-momentum tensor is conserved and Rastall gravity falls back to the general relativity.
Moreover, the corresponding field equation can be written as
\begin{equation}\label{N2}
{R_{\mu \nu }} + \left(\kappa \lambda-\frac12\right) {g_{\mu \nu }}R = \kappa {T_{\mu \nu }} ,
\end{equation}
where ${\kappa}$ is a constant related to the trace anomaly of the energy momentum tensor in Rastall gravity.

In this paper, we consider a spherically symmetric black hole solution surrounded by quintessence fluid and power-Maxwell field first derived in~\cite{grg-higher-power-maxwell-bh}.
To be specific, the metric of a static, spherically symmetric four-dimensional spacetime reads
\begin{equation}\label{N3}
d{s^2} =  - f(r)d{t^2} + f{(r)^{ - 1}}d{r^2} + {r^2}d{\theta ^2} + {r^2}{\sin ^2}\theta d{\phi ^2}.
\end{equation}
The action for the electromagnetic sector is of nonlinear form, which can be written as~\cite{epjc-charged-bh}
\begin{equation}\label{N4}
{{\cal L}_F} =  - {( - \xi {\cal F})^s},
\end{equation}
where ${\cal F} = {F_{\mu \nu }}{F^{\mu v}}{\rm{ }}$ and ${F_{\mu \nu }}$ is Faraday tensor, and $s$ and $\xi $ are constants.
If $s = 1$, it corresponds to the case of linear electromagnetic field.
Due to the presence of the electromagnetic field and quintessence fluid, the resulting energy momentum tensor possesses the form
\begin{equation}\label{N5}
T_v^\mu  = E_v^\mu  + T_v^{*\mu },
\end{equation}
where the electromagnetic part, $E_v^\mu$, is given by
\begin{equation}\label{N6}
E_\mu ^v =  - {( - \xi )^s}{({\cal F})^{s - 1}}\left( {2s{F_{\sigma \mu }}{F^{\sigma v}} - \frac{1}{2}\delta _\mu ^v{\cal F}} \right).
\end{equation}
$T_v^{*\mu }$ is the energy-momentum tensor of the quintessence field.
If one assumes a barotropic equation of state $p = w\rho $, it can be written as
\begin{equation}\label{N7}
T_t^{*t} = T_r^{*r} =  - \rho (r),T_\theta ^{*\theta } = T_\varphi ^{*\varphi } = \frac{1}{2}\rho (r)(3w + 1).
\end{equation}
By substituting the energy momentum tensor into the field equation Eq.~({\ref{N2}}), it can be shown that the black hole metric reads~\cite{grg-higher-power-maxwell-bh}
\begin{equation}\label{N10}
\begin{array}{l}
f(r) = 1 - \frac{{2M}}{r} + \frac{{{Q^2}{r^{\frac{2}{{1 - 2s}}}}{{( - 1 + 2s)}^{\frac{{3 - 2s}}{{1 - 2s}}}}}}{{(3 - 2s)s}} + \frac{{{C_a}{r^{\frac{{1 - 6\kappa \lambda  + w (3 - 6\kappa \lambda )}}{{ - 1 + 3(1 + w)\kappa \lambda }}}} \times {{(1 - 3(1 + w)\kappa \lambda )}^2}}}{{3( - 1 + 4\kappa \lambda )(\kappa \lambda  + w( - 1 + \kappa \lambda ))}} ,
\end{array}
\end{equation}
where ${C_a}$ is a constant of integration and $M$ is related to the mass of the black hole.

For the present study, we will focus on the case of the linear electromagnetic field with $s=1$ and the sole presence of a constant magnetic field, namely, the electric charge vanishes $Q=0$.
When black hole is in the magnetic field, as a good approximation\footnote{We delegate the validity of this approximation to the Appendix.}, the vector potential of the electromagnetic field can be written as
\begin{equation}\label{N11}
{A_\mu } = \frac{1}{2}B{r^2}{\sin ^2}\theta (0,0,0,1).
\end{equation}

\section{Quasinormal modes of charged massless scalar perturbations} \lb{section3}

The temporal evolution of a massless charged scalar field is governed by the Klein-Gordon equation, as follows
\begin{equation}\label{N12}
{g^{\mu \nu }}\left( {{\nabla _\mu } - iq{A_\mu }} \right)\left( {{\nabla _\nu } - iq{A_\nu }} \right)\Phi  = 0 .
\end{equation}
In order to simplify the above equation, some reasonable assumptions have to be introduced~\cite{prd-qnm-magn-bh}.
First, one has ${\nabla _\alpha }{A^\alpha } = 0$ by choosing the Lorentz gauge.
Also, one assumes that ${q^2}{B^2} \ll 1$ by considering the scenario where the coupling between the scalar field and the electromagnetic field is sufficiently weak.
Under these conditions, Eq.~(\ref{N12}) can be rewritten as
\begin{equation}\label{N13}
\frac{1}{{{r^2}}}\frac{\partial }{{\partial r}}\left( {{r^2}f\frac{{\partial \Phi }}{{\partial r}}} \right) - \left( {\frac{{{L^2}}}{{{r^2}}} - qB{L_z}} \right)\Phi  - \frac{1}{f}\frac{{{\partial ^2}\Phi }}{{\partial {t^2}}} = 0 ,
\end{equation}
where
\begin{equation}\label{N14}
{{L^2} =  - \left[ {\frac{1}{{\sin \theta }}\frac{\partial }{{\partial \theta }}\left( {\sin \theta \frac{\partial }{{\partial \theta }}} \right) + \frac{1}{{{{\sin }^2}\theta }}\frac{{{\partial ^2}}}{{\partial {\phi ^2}}}} \right]} 
\end{equation}
and
\begin{equation}\label{N15}
{{L_z} =  - i\frac{\partial }{{\partial \phi }}} .
\end{equation}

Now one may employ the method of separation of variables by writing the wave function as
\begin{equation}\label{N16}
\Phi (t,r,\theta ,\phi ) = \frac{1}{{2\pi }}\int d \omega {e^{ - i\omega t}}\sum\limits_l {\frac{{{R_{lm}}(r,\omega )}}{r}} {Y_{lm}}(\theta ,\phi ) ,
\end{equation}
where ${Y_{\ell m}}(\theta ,\phi )$ are the spherical harmonics and the radial part of the wave function ${R_{\ell m}}(r,\omega )$ satisfies 
\begin{equation}\label{N19}
\left( {\frac{{{d^2}}}{{d{r_*}^2}} + {\omega ^2} - {V_{eff}}(r)} \right){R_{lm}}(r,\omega ) = 0 ,
\end{equation}
where the effective potential is given by
\begin{equation}\label{N20}
{V_{eff}}(r) = f(r)\left[ {\frac{{l(l + 1)}}{{{r^2}}} + \frac{{f'(r)}}{r} - mqB} \right],
\end{equation}
and ${r_*}$ is the tortoise coordinate defined by $d{r_*} = \frac{{dr}}{{f(r)}}$.
One note that ${V_{eff{\rm{ }}}} =  - mqB $ as $r \to \infty $.

\begin{table}[]
\caption{
The QNMs of the massless charged scalar field in the magnetized black hole metric in Rastall gravity.
The sixth order WKB approximation and matrix method have been employed for the calculations, by assuming the metric given by Eq.~(\ref{N21}).}
\setlength{\tabcolsep}{0.3mm}
\begin{tabular}{|c|c|cc|cc|cc|cc|}
\hline
\multicolumn{1}{|c|}{\multirow{2}{*}{$l$}} & \multicolumn{1}{c|}{\multirow{2}{*}{$m$}} & \multicolumn{2}{c|}{$qB = 0$}                                        & \multicolumn{2}{c|}{$qB = 0.05$}                                        & \multicolumn{2}{c|}{$qB = 0.1$}                                        & \multicolumn{2}{c|}{$qB = 0.15$}                                         \\ \cline{3-10} 
\multicolumn{1}{|c|}{}                   & \multicolumn{1}{c|}{}                   & \multicolumn{1}{c|}{WKB}            & \multicolumn{1}{c|}{Matrix method} & \multicolumn{1}{c|}{WKB}            & \multicolumn{1}{c|}{Matrix method} & \multicolumn{1}{c|}{WKB}            & \multicolumn{1}{c|}{Matrix method} & \multicolumn{1}{c|}{WKB}             & \multicolumn{1}{c|}{Matrix method} \\ \hline
\multirow{3}{*}{1}                       & 1                                       & \multicolumn{1}{c|}{0.298-0.096i} & 0.298-0.096i          & \multicolumn{1}{c|}{0.277-0.108i} & 0.281-0.109i          & \multicolumn{1}{c|}{0.256-0.120i} & 0.266-0.127i          & \multicolumn{1}{c|}{0.236-0.130 i} & 0.200-0.178i          \\ \cline{2-10} 
                                         & 0                                       & \multicolumn{1}{c|}{0.298-0.096i} & 0.298-0.096i          & \multicolumn{1}{c|}{0.298-0.096i} & 0.298-0.096i          & \multicolumn{1}{c|}{0.298-0.096i} & 0.298-0.096i          & \multicolumn{1}{c|}{0.298-0.096 i} & 0.298-0.096i          \\ \cline{2-10} 
                                         & -1                                      & \multicolumn{1}{c|}{0.298-0.096i} & 0.298-0.096i          & \multicolumn{1}{c|}{0.320-0.083i} & 0.320-0.085i          & \multicolumn{1}{c|}{0.342-0.069i} & 0.343-0.069i          & \multicolumn{1}{c|}{0.363-0.052 i} & 0.363-0.053i          \\ \hline
\multirow{5}{*}{2}                       & 2                                       & \multicolumn{1}{c|}{0.492-0.095i} & 0.492-0.095i          & \multicolumn{1}{c|}{0.462-0.105i} & 0.464-0.104i          & \multicolumn{1}{c|}{0.433-0.115i} & 0.438-0.115i          & \multicolumn{1}{c|}{0.404-0.124 i} & 0.438-0.115i          \\ \cline{2-10} 
                                         & 1                                       & \multicolumn{1}{c|}{0.492-0.095i} & 0.492-0.095i          & \multicolumn{1}{c|}{0.477-0.100i} & 0.477-0.100i          & \multicolumn{1}{c|}{0.462-0.105i} & 0.464-0.104i          & \multicolumn{1}{c|}{0.448-0.111 i} & 0.451-0.108i          \\ \cline{2-10} 
                                         & 0                                       & \multicolumn{1}{c|}{0.492-0.095i} & 0.492-0.095i          & \multicolumn{1}{c|}{0.492-0.095i} & 0.492-0.095i          & \multicolumn{1}{c|}{0.492-0.095i} & 0.492-0.095i          & \multicolumn{1}{c|}{0.492-0.095 i} & 0.492-0.095i          \\ \cline{2-10} 
                                         & -1                                      & \multicolumn{1}{c|}{0.492-0.095i} & 0.492-0.095i          & \multicolumn{1}{c|}{0.508-0.090i} & 0.508-0.091i          & \multicolumn{1}{c|}{0.523-0.085i} & 0.524-0.085i          & \multicolumn{1}{c|}{0.539-0.080 i} & 0.539-0.079i          \\ \cline{2-10} 
                                         & -2                                      & \multicolumn{1}{c|}{0.492-0.095i} & 0.492-0.095i          & \multicolumn{1}{c|}{0.523-0.085i} & 0.524-0.085i          & \multicolumn{1}{c|}{0.555-0.074i} & 0.555-0.073i          & \multicolumn{1}{c|}{0.587-0.062 i} & 0.587-0.063i          \\ \hline
\end{tabular}
\label{tab1}
\end{table}

\begin{table}[]
\caption{The QNMs of the massless charged scalar field in the magnetized black hole metric in Rastall gravity, obtained using the code of Prony method developed in~\cite{agr-qnm-lq-matrix-06}}
\setlength{\tabcolsep}{8mm}
\begin{tabular}{|c|c|c|c|c|c|}
\hline
$l$               & $m$            & $qB = 0$                        & $qB = 0.05$                       & $qB = 0.1$                              &  $qB = 0.15$                           \\ \hline
\multirow{4}{*}{1} & 1                  & 0.293-0.098i                  & 0.272-0.113i                  & 0.247-0.108i                  & 0.229-0.206i                  \\ \cline{2-6} 
                   & 0                  & 0.293-0.098i                  & 0.293-0.098i                  & 0.293-0.098i                  & 0.293-0.098i                  \\ \cline{2-6} 
                   & -1                 & 0.293-0.098i                  & 0.316-0.084i                  & 0.339-0.070i                  & 0.370-0.051i                  \\ \hline
\multirow{5}{*}{2} & 2                  & 0.484-0.097i                  & 0.453-0.108i                  & 0.427-0.120i                  & 0.377-0.114i                  \\ \cline{2-6} 
                   & 1                  & 0.484-0.097i                  & 0.468-0.102i                  & 0.453-0.108i                  & 0.438-0.113i                  \\ \cline{2-6} 
                   & 0                  & 0.484-0.097i                  & 0.484-0.097i                  & 0.484-0.097i                  & 0.484-0.097i                  \\ \cline{2-6} 
                   & -1                 & 0.484-0.097i                  & 0.500-0.091i                  & 0.516-0.086i                  & 0.532-0.080i                  \\ \cline{2-6} 
                   & -2                 & 0.484-0.097i                  & 0.516-0.086i                  & 0.548-0.074i                  & 0.581-0.061i                  \\ \hline
\end{tabular}
\label{tab2}
\end{table}

In what follows, we evaluate the QNMs by using the sixth order WKB approximation~\cite{aj-nm-semianal-wkb-01,prd-nm-wkb-02,prd-nm-wkb-03,prd-qnm-d-bh,prd-qnm-semianaly,cqg-higher-WKB} and matrix method~\cite{agr-qnm-lq-matrix-01,agr-qnm-lq-matrix-02,agr-qnm-lq-matrix-03,agr-qnm-lq-matrix-04}.
Also, the temporal evolutions of the perturbation are investigated by using the finite difference method~\cite{prd-late-time-stell,prd-qnm-lateti-Linear-02}.
One also studies how the resulting quasinormal frequencies depend on the spacetime configurations.
Moreover, one uses the Prony method~\cite{agr-qnm-16, agr-qnm-lq-matrix-06} to extract the complex frequencies and compare them to those obtained by the WKB approximation and matrix method.
The matrix method discretizes the master equation and approximates it by a matrix equation that can eventually be solved by a nonlinear equation solver.
Of course, the QNMs can also be investigated by other well-known approaches such as the continued fraction method~\cite{agr-qnm-continued-fraction-01}.
To be more specific, the parameterization given below by Eq.~\eqref{N21} is analytic, and therefore it is expected that one can derive a three-term recurrence relation and apply the method.
For the present study, however, we will content ourselves with the three methods discussed above.
Regarding the QNM frequencies, we will be interested in the role of the magnetic field.
In addition to the choice of $s=1$ and $Q=0$ as discussed above, we consider a simplied form of the metric by assuming the following metric parameters $w =  - \frac{{10}}{9}$, $\kappa  = 1$, ${C_a} =  - 9$, and $\lambda  =  - 16$.
The resultant metric reads
\begin{equation}\label{N21}
f = 1 - \frac{{2M}}{r} + \frac{3}{{10{r^3}}}.
\end{equation}
\begin{figure*}[htbp]
\centering
\includegraphics[scale=0.7]{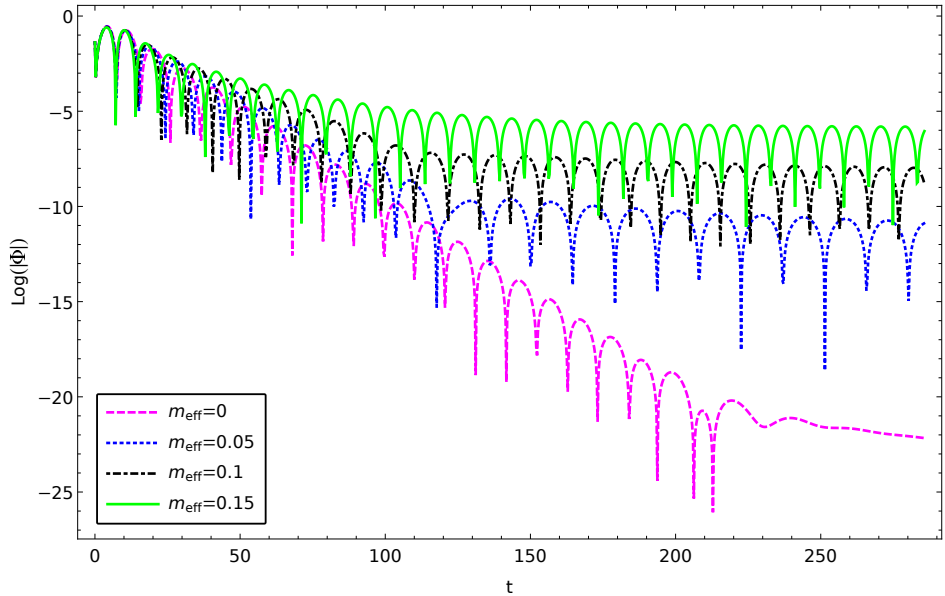}
\caption{(Color Online) The calculated temporal evolutions of charged massless scalar perturbations for different value of effective mass.
The results are obtained by using the finite difference method using the paramters $l=1$ and $M=1$.
}\label{Fig1}
\end{figure*}

\begin{figure*}[htbp]
\centering
\includegraphics[scale=0.7]{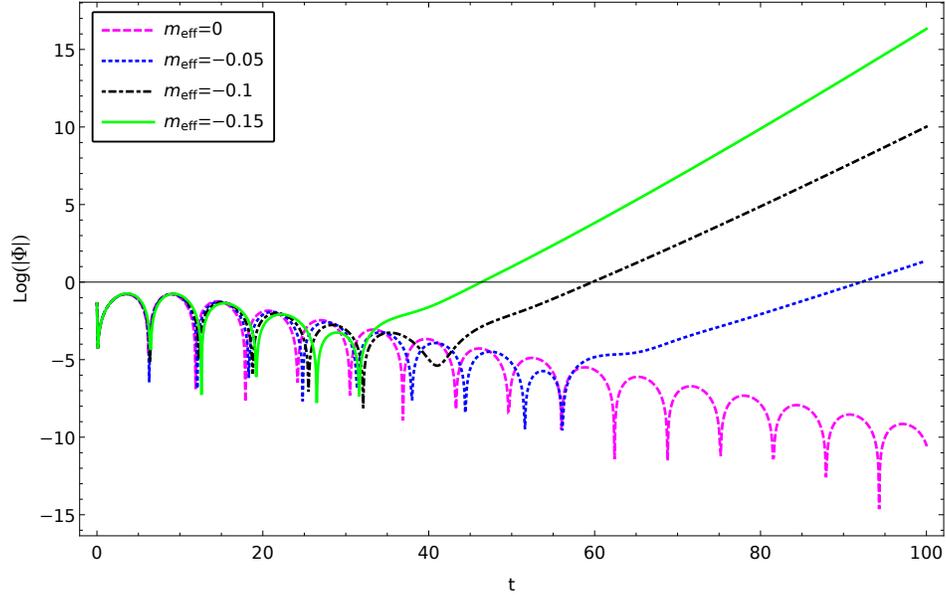}
\caption{(Color Online) The calculated temporal evolutions of charged massless scalar perturbations where instability is observed in the case of negative effective mass.
The results are obtained by using the finite difference method using the parameters $l=1$ and $M=1$.
}\label{Fig2}
\end{figure*}

\begin{figure*}[htbp]
\centering
\includegraphics[scale=0.7]{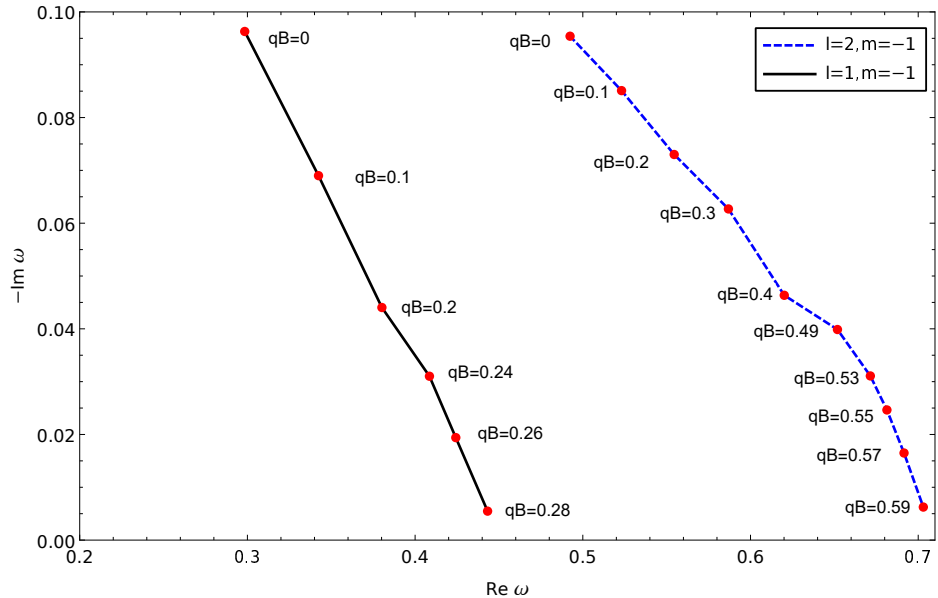}
\caption{(Color Online) The QNMs of different effective mass are calculated by matrix method for the massless charged scalar field with $l = 1,2$, $m = -1$, and $n = 0$.
}\label{Fig3}
\end{figure*}

\begin{figure*}[htbp]
\centering
\includegraphics[scale=0.7]{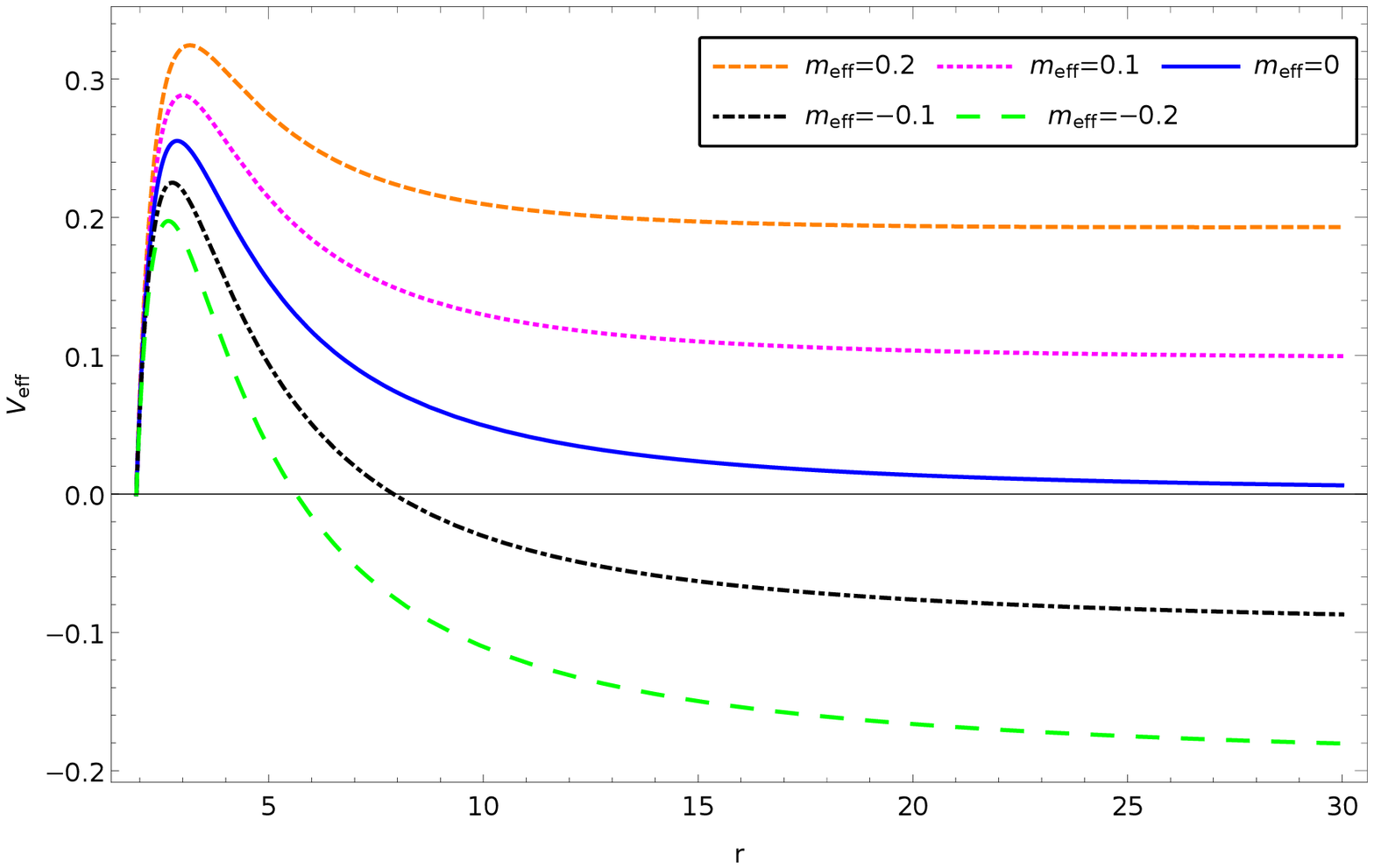}
\caption{(Color Online) The effective potentials ${V_{eff}}(r)$ for given $l = 2$ and different effective mass. 
}\label{Fig4}
\end{figure*}

The resultant quasinormal frequencies are presented in Tabs.~\ref{tab1}-\ref{tab2} and Fig.~\ref{Fig1}-\ref{Fig4}.
From Tabs.~\ref{tab1}-\ref{tab2}, one first concludes that the results obtained by three different approaches are in reasonable agreement.
In particular, the results of the matrix method and the sixth-order WKB approximation are mostly consistent.
It is observed that the term ${m_{eff}} =  - mqB$ plays the role of an effective ``mass" of the initially massless scalar field.
For positive effective mass ${m_{eff}}$, as the effective mass increases, the real part of the quasinormal frequency increases while the magnitude of the imaginary part decreases.
This indicates that the period of the dissipative oscillations becomes smaller, while the amplitude decreases slower in time.
At vanishing magnetic quantum number $m=0$, as expected, the quasinormal frequencies are independent of the strength of the magnetic field $B$.

These results are also in accordance with the temporal evolution calculated by using the finite difference method shown in Fig.~\ref{Fig1}.
In particular, as also shown in Fig.~\ref{Fig2}, as the value of ${m_{eff}}$ increases, the imaginary part will eventually become insignificant, as the envelope of the oscillations tends to lie horizontally.
It is shown explicitly in Fig.~\ref{Fig3}, where we calculate the quasinormal frequencies as functions of the effective mass using the matrix method. 
As the effective increase, the real part of the quasinormal frequency increases, and the imaginary part decreases to approach zero.
Subsequently, the resulting temporal evolutions are featured by the so-called quasi-resonance~\cite{cqg-massi-qnm,plb-decay-massi-sch}.
When one further increases the effective mass until it exceeds a critical value, the quasinormal oscillation will disappear entirely.
The latter is demonstrated in Fig.~\ref{Fig3} as the quasinormal spectrum approaches the real axis as the magnetic field increases.
As discussed in the next section, such a configuration turns out to be favorable to investigate the properties of the late-time tail in a more transparent fashion.

For negative ${m_{eff}} < 0$, the WKB and matrix method fail to produce meaningful results, and one has to resort to the finite difference method entirely.
This is signaled by the fact that the corresponding effective potential possesses a region of negative values, which indicates possible instability.
The form of the relevant effective potential, governed by Eq.~({\ref{N20}}), is shown in Fig.~\ref{Fig4}.
Similar to Ref.~\cite{prd-qnm-magn-bh}, one finds that such a configuration indeed leads to instability by exploring the temporal evolution of the initial perturbations.
To further verify this instability, we use the finite difference method to give the temporal evolution of the massless charged scalar perturbations.
As demonstrated in Fig.~\ref{Fig2} as one gradually decreases the value of the effective mass, the oscillation eventually becomes unstable.
Moreover, it is worth pointing out that there is a distinction between the quasinormal oscillation and the late-time tail in the limit when the magnetic field vanishes, which will be elaborated further in the next section.

\section{Late-time tails of charged massless scalar perturbations} \lb{section4}

In this section, we proceed to explore the late-time tails of the charged massless scalar perturbations using the analytic method proposed in Refs.~\cite{prd-decomp-perturb} as well as the numerical integration.

In terms of the Green's function, the wave function of the scalar field can be given by~\cite{book-methods-mathematical-physics-04}
\begin{equation}\label{N22}
\psi \left( {{r_*},t} \right) = \int {\left[ {G\left( {{r_*},{r_*}^\prime ;t} \right){\psi _t}\left( {{r_*}^\prime ,0} \right) + {G_t}\left( {{r_*},{r_*}^\prime ;t} \right)\psi \left( {{r_*}^\prime ,0} \right)} \right]} d{r_*}^\prime ,
\end{equation}
where the initial conditions are encoded in ${\psi \left( {{r_*} ,0} \right)}$ and $\psi_t\equiv {\partial _t}\psi ({r_*} ,0)$, and the Green's function ${G\left( {{r_*},{r_*}^\prime ;t} \right)}$ is defined by
\begin{equation}\label{N23}
\left[ {\frac{{{\partial ^2}}}{{\partial {t^2}}} - \frac{{{\partial ^2}}}{{\partial r_*^2}} + V} \right]G\left( {{r_*},{r_*}^\prime ;t} \right) = \delta (t)\delta \left( {{r_*} - {r_*}^\prime } \right),
\end{equation}
Once $\tilde G\left( {{r_*},{r_*}^\prime ;\omega } \right)$ is obtained, the evolution of an arbitrary initial perturbation can be obtained through the integration given by Eq.~({\ref{N22}}).
Indeed, both the QNM and late-time tail can be understood in terms of the singularities and branch cuts of the Fourier transform of the Green's function.
\begin{equation}\label{N24}
\tilde G\left( {{r_*},{r_*}^\prime ;\omega } \right) = \int_{{0^ - }}^{ + \infty } G \left( {{r_*},{r_*}^\prime ;t} \right){e^{i\omega t}}dt, 
\end{equation}
whose inverse transform is given by
\begin{equation}\label{N25}
G\left( {{r_*},{r_*}^\prime ;t} \right) =  - \frac{1}{{2\pi }}\int_{ - \infty  + ic}^{\infty  + ic} {\tilde G} \left( {{r_*},{r_*}^\prime ;\omega } \right){e^{ - i\omega t}}d\omega ,
\end{equation}
where $c$ is a positive constant.
Therefore, the analysis of the properties of Green's function turns into an essential task.

It is well-known that Green's function $\tilde G\left( {{r_*},{r_*}^\prime ;\omega } \right)$ can be constructed by two independent solutions of the corresponding homogeneous equation
\begin{equation}\label{N26}
\left( {\frac{{{d^2}}}{{dr_*^2}} + {\omega ^2} - V} \right){\tilde \psi _i} = 0 .
\end{equation}
These two solutions, ${{{\tilde \psi }_1}\left( {{r_*} ,\omega } \right)}$ and ${{{\tilde \psi }_2}\left( {{r_*},\omega } \right)}$, satisfy the appropriate boundary condition at both spatial boundaries~\cite{agr-qnm-12}, namely,
\begin{equation}\label{N28}
\begin{array}{l}
{{\tilde \psi }_1}\left( {{r_*},\omega } \right) \sim \left\{ {\begin{array}{*{20}{l}}
{{e^{ - i\omega {r_*}}}}&{{r_*} \to  - \infty }\\
{A(\omega ){e^{i\omega {r_*}}} + B(\omega ){e^{ - i\omega {r_* }}}}&{{r_*} \to  + \infty }
\end{array}} \right. ,
\end{array}
\end{equation}
\begin{equation}\label{N29}
\begin{array}{l}
{{\tilde \psi }_2}\left( {{r_*},\omega } \right) \sim \left\{ {\begin{array}{*{20}{l}}
{C(\omega ){e^{i\omega {r_*}}} + D(\omega ){e^{ - i\omega {r_*}}}}&{{r_*} \to  - \infty }\\
{{e^{ + i\omega {r_*}}}}&{{r_*} \to  + \infty }
\end{array}} \right. .
\end{array}
\end{equation}

It is not difficult to show that the following form satisfies Eq.~({\ref{N23}})
\begin{equation}\label{N27}
\begin{array}{l}
\tilde G\left( {{r_*},{r_*}^\prime ;\omega } \right) =  - \frac{1}{{W(\omega )}}\left\{ {\begin{array}{*{20}{l}}
{{{\tilde \psi }_1}\left( {{r_*}^\prime ,\omega } \right){{\tilde \psi }_2}\left( {{r_*},\omega } \right)}&{,\quad {r_*}^\prime  > {r_*}}\\
{{{\tilde \psi }_1}\left( {{r_*},\omega } \right){{\tilde \psi }_2}\left( {{r_*}^\prime ,\omega } \right)}&{,\quad {r_*}^\prime  < {r_*}},
\end{array}} \right.
\end{array}
\end{equation}
where 
\begin{equation}\label{WronsDef}
W(\omega) \equiv {\tilde \psi }_1\left( {{r_*} ,\omega } \right){\tilde \psi }_{2,r_*}\left( {{r_*} ,\omega }\right) - {\tilde \psi }_2\left( {{r_*} ,\omega } \right) {\tilde \psi }_{1,r_*}\left( {{r_*} ,\omega } \right)
\end{equation}
is the Wronskian of ${\tilde \psi }_1\left( {{r_*} ,\omega } \right)$ and ${\tilde \psi }_2\left( {{r_*} ,\omega } \right)$.

It is understood~\cite{prd-nonspherical-scalar-gravita} that the late-time tails of quasinormal oscillations can be attributed to the branch cut of the Green's function.
For the massless scalar field, the branch cut lies on the negative part of the imaginary axis~\cite{agr-qnm-tail-06}.
For the massive scalar field, on the other hand, Hod and Piran were the first to point out~\cite{prd-tail-gravi-collap} that the branch cut is lying on the real axis, between the two branch points governed by the mass.
Using reasonable approximation, an appropriate estimation of the contribution from the branch cut subsequently gives rise to the main feature of the late-time tail.
From a physical viewpoint, since the low-frequency waveforms are more likely to be backscattered by the curvature at the asymptotic infinity, the late-time tail is therefore primarily dominated by the low-frequency contributions~\cite{prd-tail-gravi-collap}.
Mathematically, on the other hand, this is because both types of late-time tails are originated from the branch cut, which stays closer to the real axis than any QNM pole.

To proceed, we introduce
\begin{equation}\label{N30}
{R_{lm}}(r,\omega ) = {\left( {1 - \frac{{2M}}{r} + {r^{ - (2 + a)}}} \right)^{ - 1/2}}\tilde \psi .
\end{equation}
By further following Refs.~\cite{prd-tail-gravi-collap, Koyama2001, prd-monopole, prd-massi-sch}, one assumes that the initial perturbation and the observer are both located away from the black hole.
Subsequently, the master equation can be simplified by expanding Eq.~(\ref{N19}) in $\frac{M}{r}$\footnote{This approximation is also justified by the consistent results obtained from straightforward numerical calculations.}.
To be specific, one ignores the terms of order $\frac{c}{{{r^{\alpha }}}}$ with $\alpha  > 2$, and finds
\begin{equation}\label{N31}
\left[ {\frac{{{d^2}}}{{d{r^2}}} + {\omega ^2} + Bqm + \frac{{4M{\omega ^2} + Bqm2M}}{r} - \frac{{l\left( {l + 1} \right)}}{{{r^2}}}} \right]\tilde \psi  = 0.
\end{equation}

By further introducing
\begin{eqnarray}\label{N32}
\tilde \psi  &=& {z^{l + 1}}{e^{ - \frac{z}{2}}}\Phi (z), \nonumber\\
z &=& 2\sqrt { - Bqm - {\omega ^2}} r \equiv 2\varpi r ,\nonumber\\
\lambda  &=& \frac{{M( - Bqm)}}{\varpi } - 2M\varpi ,
\end{eqnarray}
one can write down two relevant solutions for homogeneous equation Eq.~(\ref{N26}) as follows
\begin{eqnarray}\label{N33}
{{\tilde \psi }_1} &=& A' {M_{\lambda ,(l + \frac{1}{2})}}(2\varpi r)  ,\nonumber\\
{{\tilde \psi }_2} &=& B'{W_{\lambda ,(l + \frac{1}{2})}}(2\varpi r)  ,
\end{eqnarray}
where $A' = C' \varpi^{-\left(l+1\right)} $ and $B', C'$ are some one-valued even functions of $\varpi$.
$M_{k, m}$ and $W_{k, m}$ are the Whittaker functions, which can be expressed in terms of the solutions of Kummer's equation $M(a,b,z)$ and $U(a,b,z)$, namely, the confluent hypergeometric functions~\cite{book-methods-mathematical-physics-06},
\begin{eqnarray}\label{N34}
{M_{\lambda ,(l + \frac{1}{2})}}(2\varpi r) &=& {e^{ - \varpi r}}{(2\varpi r)^{(l + \frac{1}{2}) + \frac{1}{2}}}M\left( {l + 1 - \lambda ,2l + 2,2\varpi r} \right) ,\nonumber\\
{W_{\lambda ,(l + \frac{1}{2})}}(2\varpi r) &=& {e^{ - \varpi r}}{(2\varpi r)^{(l + \frac{1}{2}) + \frac{1}{2}}}U\left( {l + 1 - \lambda ,2l + 2,2\varpi r} \right) .
\end{eqnarray}

\begin{figure*}[htbp]
\centering
\includegraphics[scale=0.4]{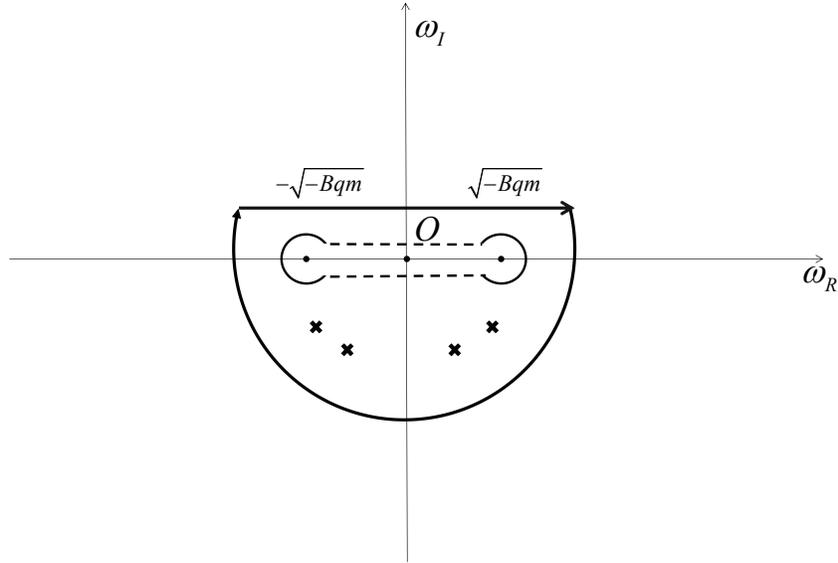}
\caption{Revelant contour of the integration on the complex frequency plane of the Green's function.
The inverse Fourier transform corresponds to the integral along the upper side of the real axis.
It can be complemented by an integral along a large semicircle in the lower half of the complex plane, which is subsequently deformed to go around the quasinormal poles and brach cuts indicated in the plot.}\label{Fig5}
\end{figure*}

It is noted that the specific choice of $C' $ guarantees the proper in-going wave boundary condition at the horizon.
Moreover, as discussed below, ${\tilde \psi }_1$ satisfies Eq.~\eqref{N51}, as it does not contain any discontinuity when crossing the branch cut~\cite{Koyama2001, prd-monopole, prd-massi-sch}.
On the other hand, ${\tilde \psi }_2$ gives rise to the outgoing wave at spatial infinity that possesses two branch points and, subsequently, a branch cut joining them.
As first proposed in~\cite{prd-tail-gravi-collap}, one may conveniently place the branching cut on the real axis of the $\omega $-plane, as illustrated in Fig.~\ref{Fig5}.
In this context, the late-time tail is mainly governed by the properties of ${\tilde \psi }_2$.
It is noted that it is mandatory to show that the branch cut is in the lower half of the complex plane.
Moreover, for the result to be physically meaningful, the specific choice of the branch cut's location must also be irrelevant.
A detailed account of this point was discussed recently in Ref.~\cite{agr-qnm-tail-39}.

By using the explicit forms of Eqs.~(\ref{N33}-\ref{N34}), one can evaluate the contribution from the difference between the two sides of the branch cut.
Therefore, the inverse Fourier transform Eqs.~(\ref{N25}) gives
\begin{equation}\label{N35}
{G^C}\left( {{r_*},r_*^\prime ;t} \right)  = \frac{1}{{2\pi }}\int_{ - \sqrt { - Bqm} }^{\sqrt { - Bqm} } F (\varpi ){e^{ - i\omega t}}d\omega ,
\end{equation}
where
\begin{equation}
F (\varpi )\equiv  {\frac{{{{\tilde \psi }_1}\left( {r_*^\prime ,\varpi {e^{i\pi }}} \right){{\tilde \psi }_2}\left( {{r_*},\varpi {e^{i\pi }}} \right)}}{{W\left( {\varpi {e^{i\pi }}} \right)}} - \frac{{{{\tilde \psi }_1}\left( {r_*^\prime ,\varpi } \right){{\tilde \psi }_2}\left( {{r_*},\varpi } \right)}}{{W(\varpi )}}}  .
\end{equation}

Regarding the integrand in Eq.~(\ref{N35}), we first calculate the Wronskian in the denominator utilizing the following relations~\cite{book-methods-mathematical-physics-06}
\begin{equation}\label{N36}
{W_{\lambda ,l + \frac{1}{2}}}(2\varpi r) = \frac{{\Gamma ( - 2l - 1)}}{{\Gamma ( - l - \lambda )}}{M_{\lambda ,l + \frac{1}{2}}}(2\varpi r) + \frac{{\Gamma (2l + 1)}}{{\Gamma (l + 1 - \lambda )}}{M_{\lambda , - \left( {l + \frac{1}{2}} \right)}}(2\varpi r) .
\end{equation}
and the non-vanishing Wronskian between the confluent hypergeometric functions
\begin{eqnarray}\label{Wrons}
W\left\{ {{M_{\lambda ,l + \frac{1}{2}}}(2\varpi r),{M_{\lambda , - \left( {l + \frac{1}{2}} \right)}}(2\varpi r)} \right\} =  - (2l + 1)(2\varpi ).
\end{eqnarray}
It is found that
\begin{equation}\label{NWrons}
W(\varpi ) = {A^\prime }{B^\prime }( - 2l - 1)(2\varpi )\frac{{\Gamma (2l + 1)}}{{\Gamma (l + 1 - \lambda )}}.
\end{equation}

To give an account for the discontinuity of the integrand across the branch cut, one makes use of 
\begin{eqnarray}\label{N53}
{M_{\pm\lambda ,l + \frac{1}{2}}}\left( {{e^{i\pi }}2\varpi r} \right) = {e^{(l + 1)i\pi }}{M_{ \mp \lambda ,l + \frac{1}{2}}}(2\varpi r), 
\end{eqnarray}
to evaluate ${\tilde \psi }_{1,2}\left( {r_* ,\varpi {e^{i\pi }}} \right)$ as follows
\begin{equation}\label{N51}
{\tilde \psi }_1\left( {r_* ,\varpi {e^{i\pi }}} \right) = {{\tilde \psi }_1}\left( {r_* ,\varpi } \right) ,
\end{equation}
\begin{equation}\label{N52}
{\tilde \psi }_2\left( {{r_*},\varpi {e^{i\pi }}} \right) = {B^\prime }\left[{e^{\left( {l + 1} \right)i\pi }}\frac{{\Gamma ( - 2l - 1)}}{{\Gamma \left( { - l + \lambda } \right)}}{M_{ \lambda ,l + \frac{1}{2}}}(2\varpi {r_*}) + {e^{\left( { - l} \right)i\pi }}\frac{{\Gamma (2l + 1)}}{{\Gamma \left( {l + 1 + \lambda } \right)}}{M_{ \lambda , - (l + \frac{1}{2})}}(2\varpi {r_*})\right] .
\end{equation}
In the above results, it is essential to note that the subscripts $\lambda$ of the Whittaker functions remain unchanged.
Eq.~\eqref{N51} manifestly indicates that ${{\tilde \psi }_1}$ does not possess any branch cut, as it is also a generic physical requirement~\cite{agr-qnm-tail-06, prd-tail-gravi-collap, Koyama2001}.
Also, the Wronskian gives
\begin{equation}\label{NWronsFlip}
W\left( {\varpi {e^{i\pi }}} \right) = A'{B^\prime }( - 2l - 1)(2\varpi )\frac{{\Gamma (2l + 1)}}{{\Gamma (l + 1 + \lambda )}}{e^{( - l)i\pi }}
\end{equation}
We note that in deriving the above result, one should always keep an eye on Eqs.~\eqref{N51} and~\eqref{N52}.
To be specific, the discontinuity associated with the factor $\varpi^{-(l+1)}$ embeded in $A'$ does not manifest itself, while the factor $e^{-l i\pi}$ on the r.h.s. of Eq.~\eqref{N52} appears.

By putting all the pieces together, Eqs.~(\ref{N51}),~(\ref{N52}), and~\eqref{NWronsFlip}, one finds that the integrand of Eq.~(\ref{N35}) gives
\begin{equation}\label{N38}
\begin{array}{l}
F(\varpi ) = \frac{1}{{( - 2l - 1)(2\varpi ){A^\prime }^2}}{{\tilde \psi }_1}\left( {r_*^\prime ,\varpi } \right)\tilde F(\varpi ){{\tilde \psi }_1}\left( {{r_*},\varpi } \right)
\end{array}
\end{equation}
where 
\begin{equation}\label{tildeF}
\tilde F(\varpi ) =
e^{(2l+1)i\pi}\frac{\Gamma\left(-2l-1\right)\Gamma\left(l+1+\lambda\right)}{\Gamma\left(-l+\lambda\right)\Gamma\left(2l+1\right)}
-\frac{\Gamma\left(-2l-1\right)\Gamma\left(l+1-\lambda\right)}{\Gamma\left(-l-\lambda\right)\Gamma\left(2l+1\right)}  .
\end{equation}
Here, due to Eq.~(\ref{N51}), ${\tilde \psi }_1\left( {r_*}^\prime ,\varpi  \right)$ can be readily factorized out. 
The remaining term, ${\tilde \psi }_2\left( {r_*} ,\varpi  \right)$, can be simplified by dropping out irrelevant contributions which possesses identical discontinuity of the Wronskian. 
As a result, a second ${\tilde \psi }_1\left( {r_*} ,\varpi  \right)$ can be pulled out.

For the intermediate-late time scale, which corresponds to the range $M \ll r \ll t \ll M/( - qBm{M^2})$, the frequency $\varpi  = O(\sqrt {\sqrt { - qBm} /t} )$ gives rise to a major contribution to the integral, which implies $\lambda  \ll 1$ due to Eq.~(\ref{N32}).
In this case, the term proportional to $\frac{1}{r}$ in Eqs.~(\ref{N31}), which describes the backscattering due to the curvature of asymptotic infinity, is insignificant~\cite{prd-tails-rn, Koyama2001}.
Subsequently, the inverse Fourier transform of the Green's function at the limit $\lambda \ll 1$ can be performed mostly analytically, which results in
\begin{equation}\label{N41}
\begin{array}{l}
{G^C}\left( {{r_*},r_*^\prime ;t} \right) = \frac{{\left( {1 - {e^{(2l + 1)i\pi }}} \right)\Gamma ( - 2l - 1)\Gamma (l + 1)\Gamma \left( {l + \frac{3}{2}} \right){{( - qBm)}^{\frac{{2l + 1}}{4}}}{{\left( {r_*^\prime {r_*}} \right)}^{l + 1}}{t^{ - \frac{{2l + 3}}{2}}}}}{{\pi (4l + 2){2^{ - 3l - \frac{5}{2}}}\Gamma (2l + 1)\Gamma ( - l)}}\\
 \times \cos \left[ {\sqrt { - qBm} t - \pi \left( {\frac{{2l + 3}}{4}} \right)} \right].
\end{array}
\end{equation}
In other words, one finds that at the intermediate-late time, the tail is dominated by the form
\begin{equation}
{t^{ - l - \frac{3}{2}}}\cos \left[ {\sqrt { - qBm} t - \pi \left(\frac{l}{2} + \frac{3}{4}\right)} \right] . \nonumber
\end{equation}
It oscillates while decaying by a power-law form.
Besides the magnetic field, the intermediate-late time tail depends on the multipole number $l$ and the azimuthal number $m$.
In particular, the presence of the magnetic field gives rise to an effective mass that significantly modifies the properties of the tail.
However, if one removes the magnetic field, the tail governed by taking the limit $B\to 0$ in Eq.~(\ref{N41}) does not simply fall back to its massless counterpart.
To be specific, while the synodal oscillation is suppressed, the remaining power-law exponent becomes $e^{-l-\frac32}$, which is different from the late-time tail $e^{-2l-3}$ observed for a massless scalar in Schwarzschild background.
Moreover, as discussed below, at the limit of the vanishing magnetic field, neither the power-law exponent of the late-time tail matches that of its massless counterpart. 
We postpone further discussions regarding this intriguing feature to the end of the present section.

Now, we turn to discuss the asymptotic behavior of the late-time tails.
At significant time scale $\sqrt { - qBm} t \gg 1/{(\sqrt { - qBm} M)^2}$, the relevant terms in the master equation behave differently from the case of intermediate-late time tail.
The backscattering from the spacetime curvature at asymptotic infinity can no longer be ignored for the present scenario.
In this case, since both the ${\tilde \psi }_1$ factors in Eq.~\eqref{N38} do not depend sensitively on the frequency $\omega$, we only need to focus on the term Eq.~\eqref{tildeF}.
By applying the above approximation and making use of the asymptotical forms of $\Gamma$ functions~\cite{book-methods-mathematical-physics-06}, one finds
\begin{equation}\label{N39}
\tilde F(\varpi ) \approx \frac{\Gamma\left(-2l-1\right)}{\Gamma\left(2l+1\right)}\lambda^{2l+1}\left[e^{(2l+1)i\pi} - \frac{\eta_+ e^{i\pi \lambda} + \eta_- e^{-i\pi \lambda}}{\eta_- e^{i\pi \lambda} + \eta_+ e^{-i\pi \lambda}}\right] ,
\end{equation}
where
\begin{eqnarray}\label{etagamma}
\eta_\pm  = \mp  e^{\pm i\pi l} .
\end{eqnarray}

The resulting integral of the inverse Fourier transform is somewhat complicated, but its asymptotic properties are governed by the terms related to $\lambda$. 
At small frequencies, we have $\varpi\to 0$ and $\lambda\to \infty$.
Their product, which appears in the argument of the Bessel functions, gives 
\begin{equation}
\lambda \varpi  = (\frac{{M( - Bqm)}}{\varpi } - 2M\varpi )\varpi  = M( - Bqm) + O(\varpi^2), \nonumber
\end{equation}
which is finite and varies slowly in the low-frequency region $\omega \to \sqrt{ - Bqm}_+$~\cite{prd-monopole}. 
Both the terms $\lambda^{2l+1}$ and $e^{\pm i\pi \lambda}$ of Eq.~\eqref{N39} oscillate significantly as $\lambda\to \infty$, while the latter type is even more drastic when compared to the former.
This indicates that the integral's main contribution comes from these terms.
In most cases, one may follow the arguments by Koyama and Tomimatsu~\cite{prd-tails-rn} that the main contribution of the integral can be obtained by employing the method of steepest descent.
The strategy of such a treatment is to isolate the part which oscillates significantly and then encounter the saddle point, where the oscillations evolve the slowest.
The remaing part might be tedious in form but varies moderately.
We rewrite the integrand as
\begin{equation}\label{N47}
\tilde F(\varpi )e^{-i\omega t} \approx \frac{\Gamma\left(-2l-1\right)}{\Gamma\left(2l+1\right)}\lambda^{2l+1}{e^{i\phi }}  e^{i(2\pi\lambda-\omega t)},
\end{equation}
where the phase can be defined by
\begin{equation}\label{N48}
{e^{i\phi }} =  - \frac{\eta_+  + \eta_- e^{-2i\pi \lambda}}{\eta_+ + \eta_- e^{2i\pi \lambda} } .
\end{equation}
As shown in~\cite{agr-qnm-tail-39}, the variation of the phase $\phi$ actually cancels precisely that in the term $e^{2i\pi\lambda}$.
In this case, the integral cannot be evaluated by the method of steepest descent.
Nonetheless, one may still evaluate the most dominant term of the integral and find
\begin{eqnarray}\label{asympTail}
{G^C}\left(r_*, {r'}_*; t\right)\sim  t^{-1}\sin(\mu t +\varphi) ,
\end{eqnarray}
where the phase shift $\varphi$ contains some minor dependence in $t$.
The temporal dependence of Eq.~\eqref{asympTail} implies that late-time tail evaluated regarding the contribution from the saddle point is $\sim {t^{ - 1}}$.
It is noted that the obtained exponential does not depend on either the angular momentum number or the strength of the magnetic field.

\begin{figure*}[htbp]
\centering
\includegraphics[scale=0.7]{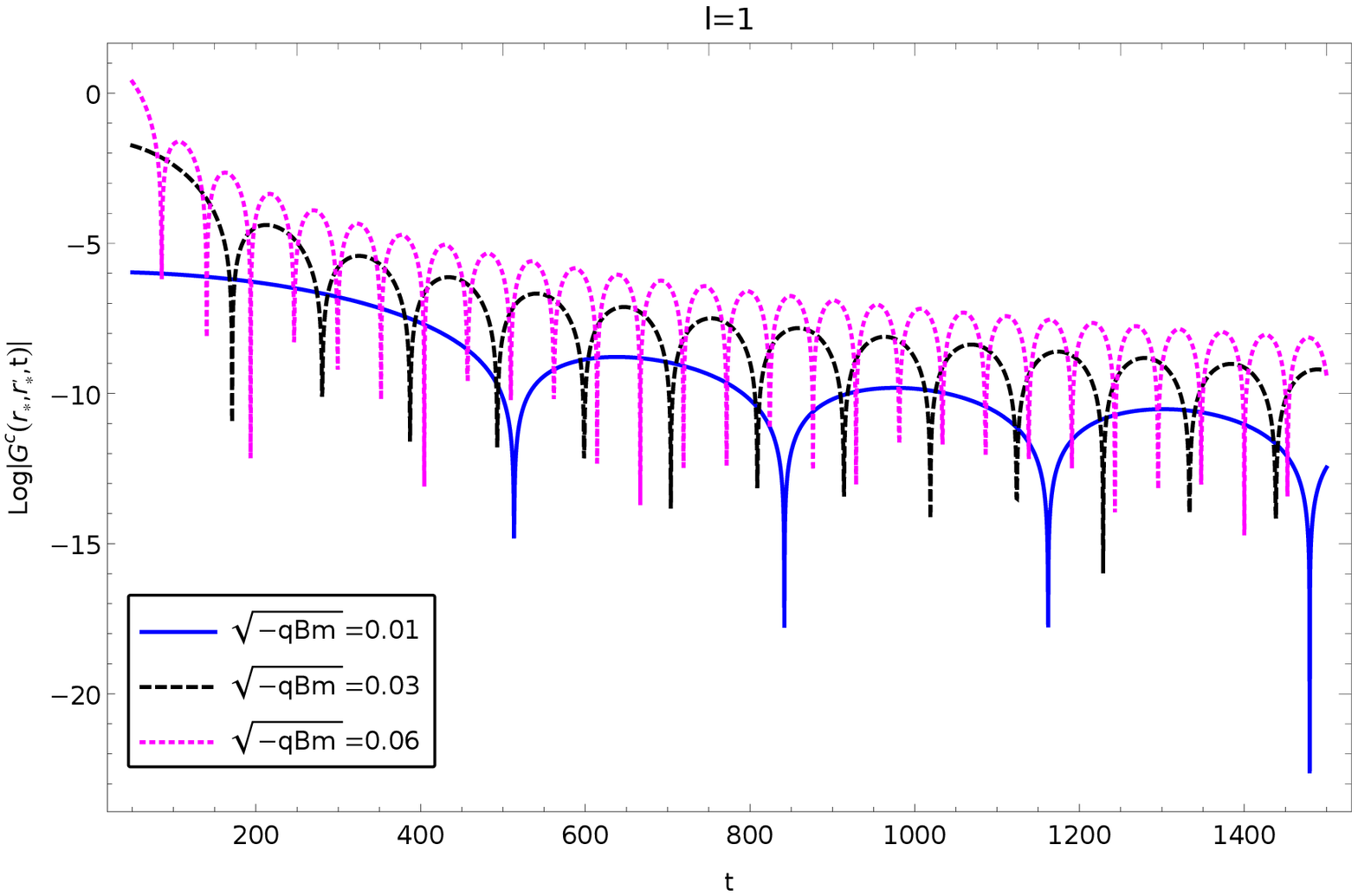}
\includegraphics[scale=0.7]{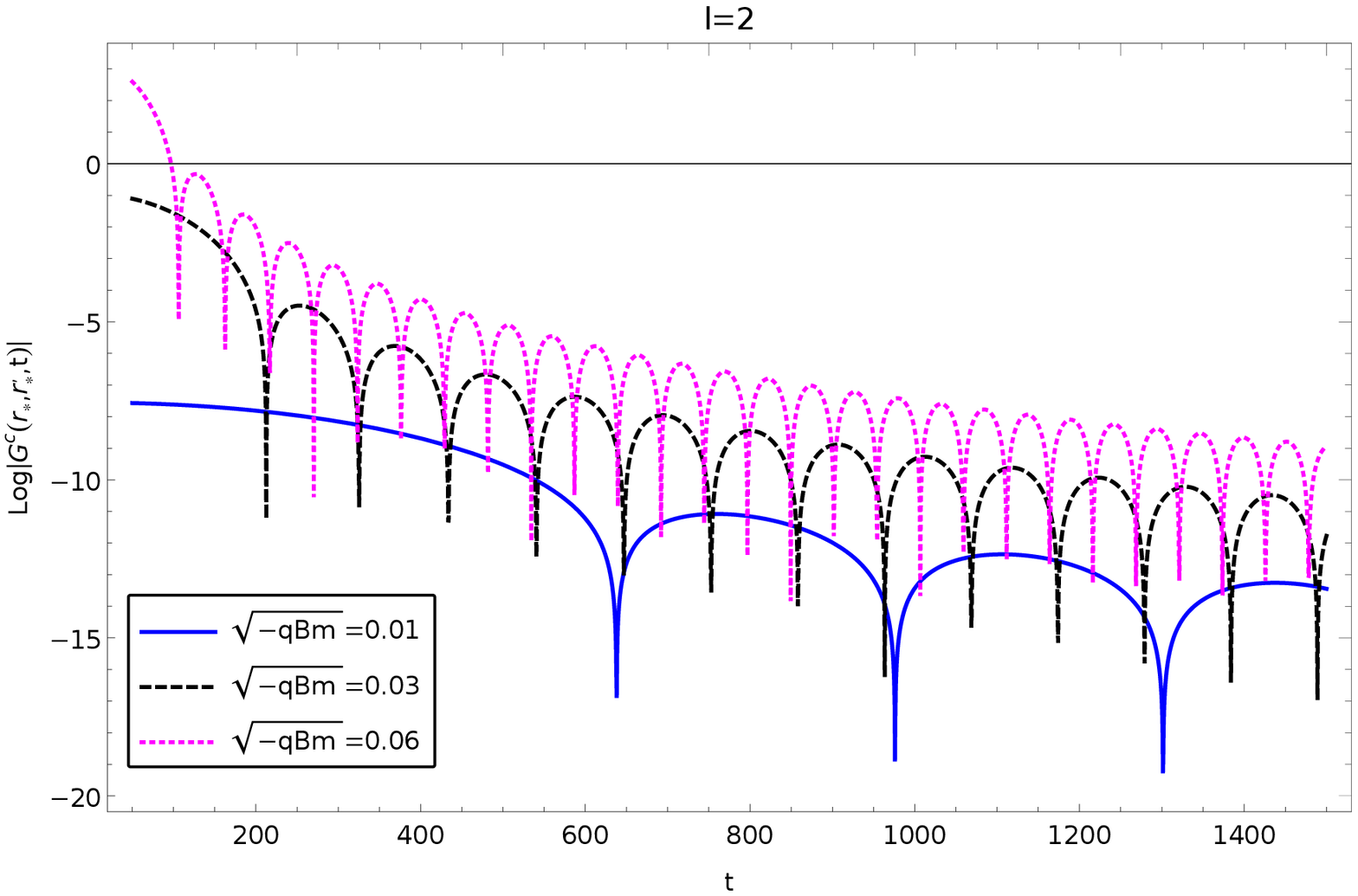}
\caption{(Color Online) The calculated the intermediate-late time tail in the Green's function of charged massless scalar perturbations in a magnetized black hole in Rastall gravity.
The results are obtained by using numerical integration for different effective masses with $l = 1, 2$.
 }\label{Fig6}
\end{figure*}

\begin{figure*}[htbp]
\centering
\includegraphics[scale=0.7]{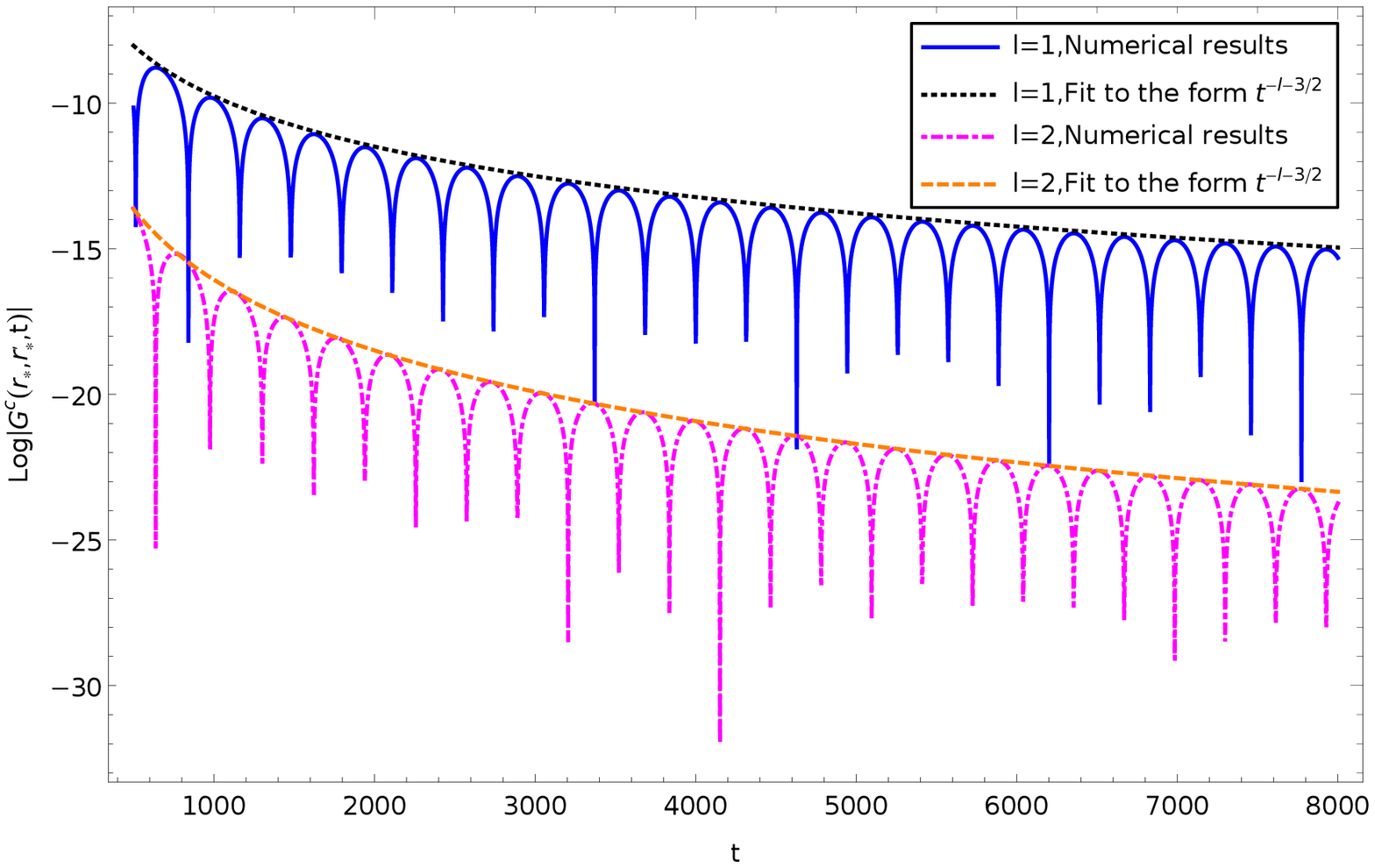}
\caption{(Color Online) The calculated the intermediate-late time tail in the Green's function of charged massless scalar perturbations in a magnetized black hole in Rastall gravity.
The results are obtained by using numerical integration for different effective masses with $l = 1, 2$.
The envelopes of dissipative oscillations shown in dashed curves are compared with the analytic ones $t^{ - (l + 3/2)}$ given in the text.
}\label{Fig7}
\end{figure*}

\begin{figure*}[htbp]
\centering
\includegraphics[scale=0.6]{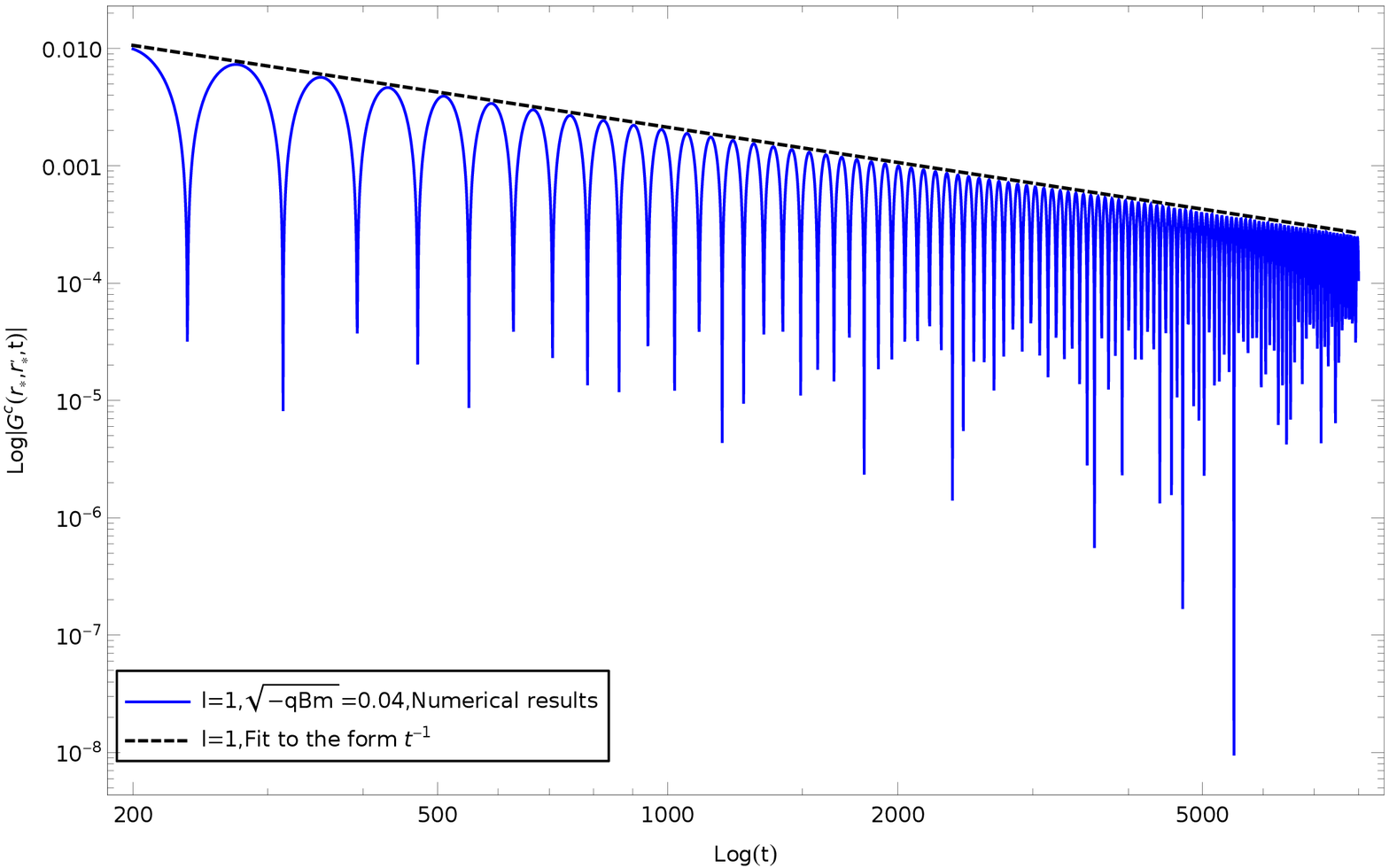}
\includegraphics[scale=0.6]{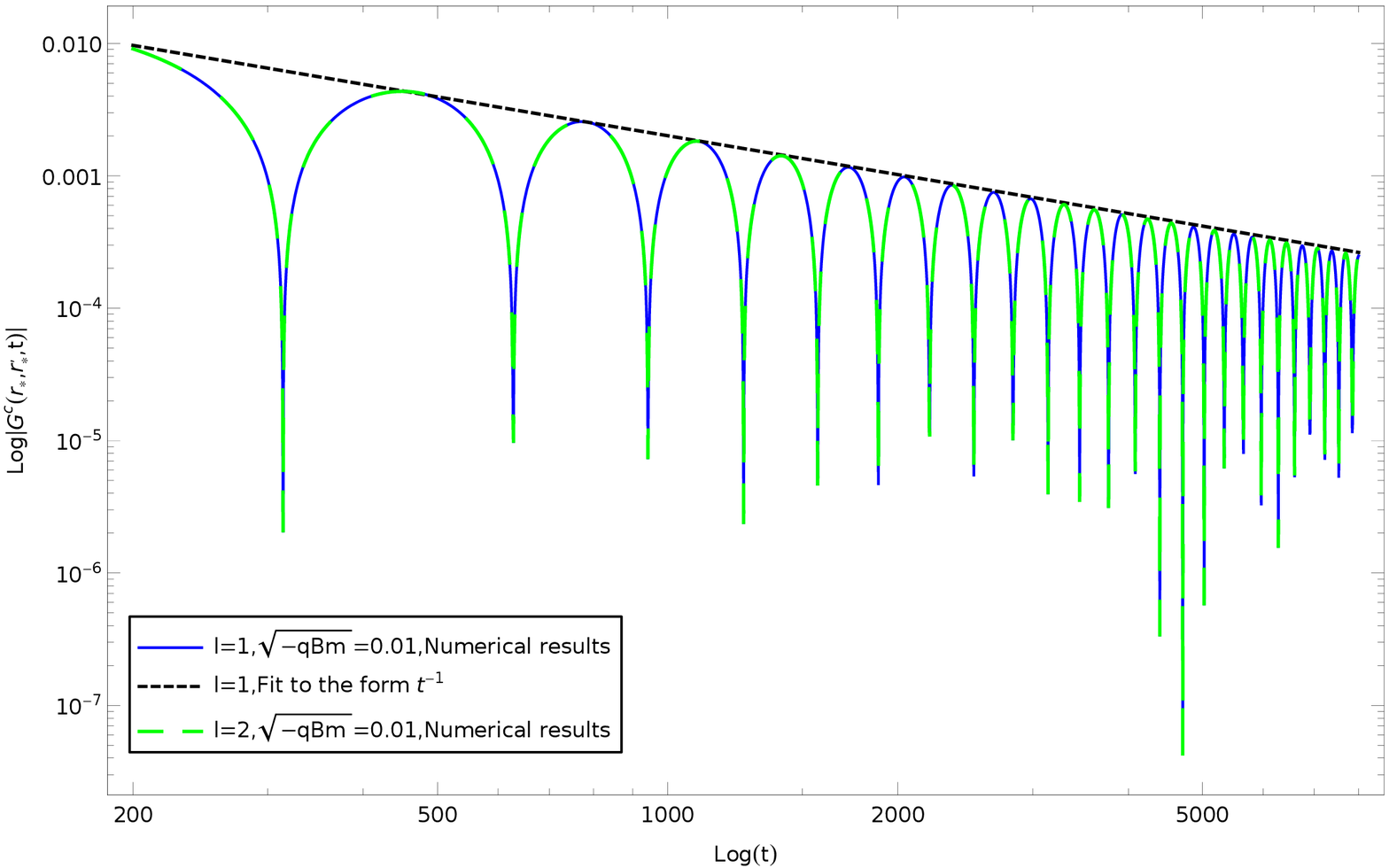}
\caption{(Color Online) The calculated the late-time tail in the Green's function of charged massless scalar perturbations in a magnetized black hole in Rastall gravity.
The results are obtained using numerical integration for different effective masses with the parameters $l = 1, 2$ and $M=1$.
The envelopes of dissipative oscillations shown in dashed curves are compared with the analytic ones $t^{ - 1}$ given in the text.
}\label{Fig8}
\end{figure*}

\begin{figure*}[htbp]
\centering
\includegraphics[scale=0.6]{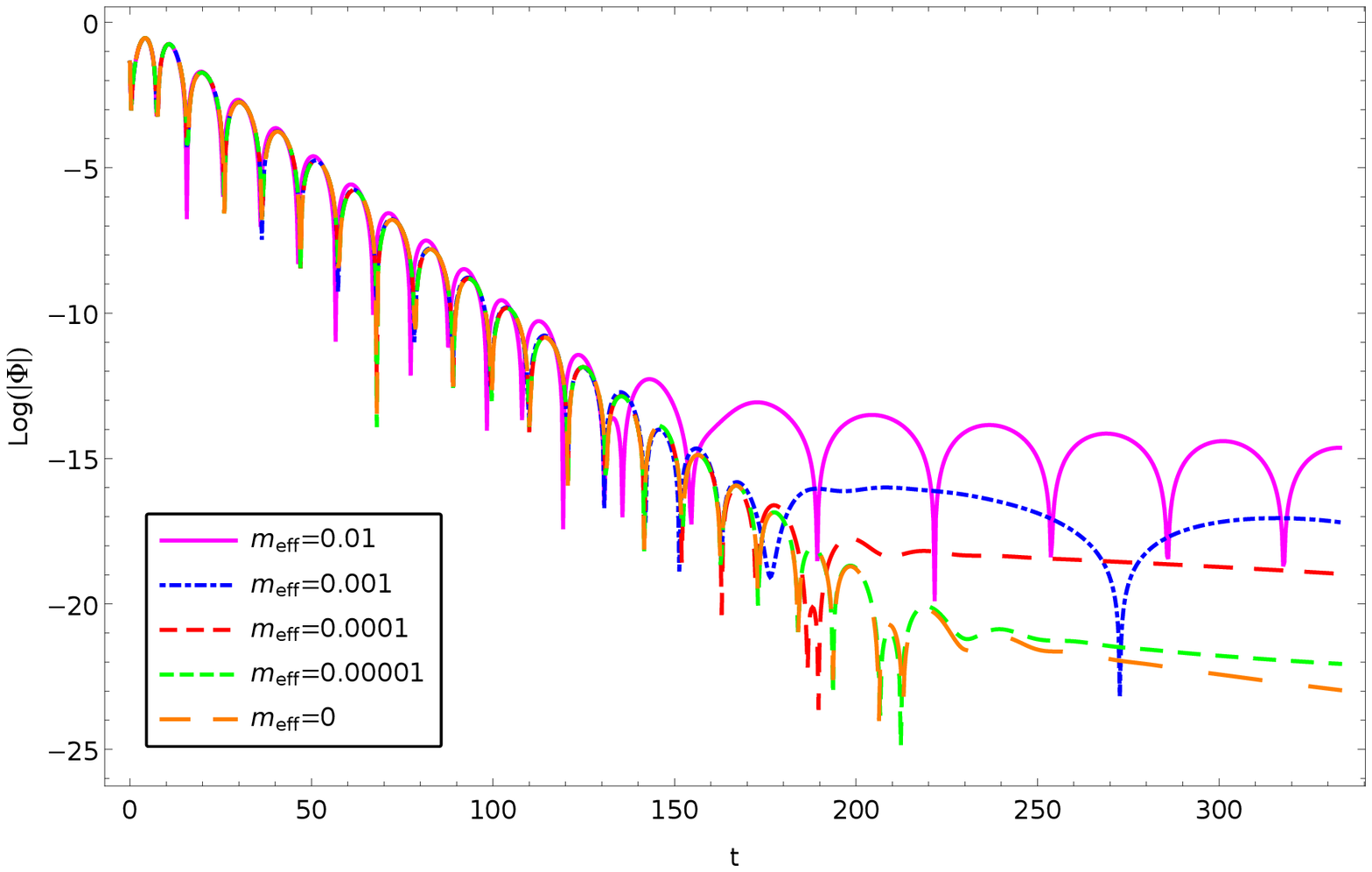}
\includegraphics[scale=0.6]{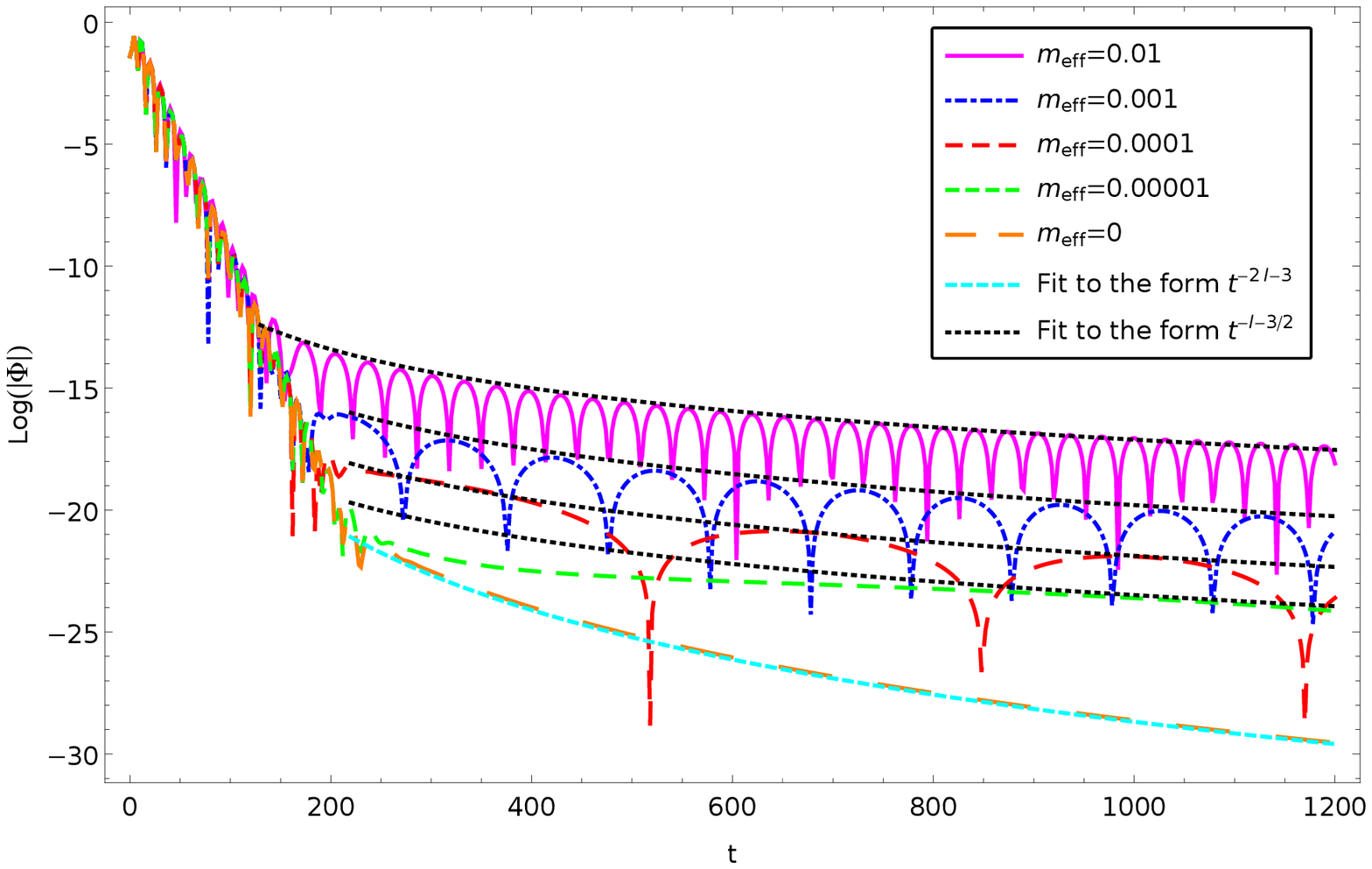}
\caption{(Color Online) The calculated the quasinormal oscillations and late-time tails of charged massless scalar perturbations in a magnetized black hole in Rastall gravity.
The calculations aim to explore the asymptotic behavior of the QNMs and tails as the effective mass gradually vanishes.
The results are obtained by using the parameters $l = 1$ and $M=1$.
Top: The quasinormal oscillations are shown to converge to the case of massless perturbations as the effective mass approaches zero.
Bottom: The envelopes of the tails are are compared with the analytic forms $t^{ - l-3/2}$ and $t^{ - 2l-3}$, respectively.
The results indicate that the tails' asymptotic behavior does not fall back to that of massless perturbations, as discussed in the text.
}\label{Fig9}
\end{figure*}

In what follows, we present the numerical results in Figs.~\ref{Fig6}-\ref{Fig9}.
In order to confirm the above analytical results, numerical calculations have been performed by using numerical integration.
The resultant temporal evolutions of the intermediate-late time are presented in Fig.~\ref{Fig6}-\ref{Fig7}.
In Fig.~\ref{Fig6}, we show the temporal evolutions at the intermediate-late time for different effective masses. 
For given $l$, following Eq.~\eqref{N41}, the period of the oscillations increases as the effective mass decreases. 
Also, it is observed in Fig.~\ref{Fig7} that when the effective mass is fixed, the oscillating frequency largely remains unchanged for different values of $l$.
On the other hand, the attenuation becomes more significant as $l$ increases.
In Fig.~\ref{Fig7}, the time profiles obtained numerically are also compared against the analytic ones shown in dashed curves given by Eq.~\eqref{N41}.
The extracted exponent from the envelope of the oscillation gives $ {t^{ - \beta }}$, where $\beta  = 2.5087 \pm 0.0020$ and $\beta  = 3.5015 \pm 0.0020$ with respect to $l = 1$, $l = 2$.
Therefore, the asymptotic form $t^{ - (l + 3/2)}$ for the intermediate-late time agrees well with the numerical results for different $l$.

The temporal evolutions of the late-time tails are presented in Fig.~\ref{Fig8}.
The calculations are carried out for different effective masses and angular momenta.
Again, by comparing the two plots in Fig.~\ref{Fig8}, as the effective mass increases, one observes that the oscillation frequency also increases, but the attenuation rate remains the same.
Moreover, as shown in the bottom plot of Fig.~\ref{Fig8}, neither the attenuation rate nor the oscillation period depends on the angular momentum $l$.
We also show in dashed curves the nonlinear fitting to the envelope of the oscillations.
The extracted exponents from the a power-law form $  {t^{ - \beta }}$ read, respectively for the two cases, $\beta  = 0.9978 \pm 0.0068$ and $\beta  = 0.9762 \pm 0.0082$, which are reasonably consistent with the analytic result $t^{-1}$.

\begin{table}[]
\caption{
The QNMs for the metric Eq.~(\ref{N21}) are calculated by reducing effective mass gradually.
The Prony method, sixth order WKB approximations and the matrix method are used for the calculations.
The results are obtained by taking $M = 1, l = 1$.}
\setlength{\tabcolsep}{8mm}
\begin{tabular}{|c|c|c|c|}
\hline
${m_{eff}}$  & Prony            & WKB              & Matrix method               \\ \hline
0       & 0.29861-0.09614i & 0.29842-0.09636i & 0.29845-0.09622i \\ \hline
0.00001 & 0.29861-0.09614i & 0.29842-0.09636i & 0.29845-0.09622i \\ \hline
0.0001  & 0.29865-0.09612i & 0.29846-0.09634i & 0.29849-0.09620i \\ \hline
0.001   & 0.29904-0.09590i & 0.29885-0.09611i & 0.29884-0.09600i \\ \hline
0.01    & 0.30288-0.09421i & 0.30276-0.09384i & 0.30246-0.09408i \\ \hline
\end{tabular}
\label{tab3}
\end{table}

Last but not least, we elaborate further on the specific role of the magnetic field in the present study.
A distinct characteristic of the present spacetime configuration is that the effective mass is generated by the magnetic field, and therefore the effect on the late-time tail is dynamic. 
In particular, we are allowed to consider a process in which one gradually tunes the magnetic field to approach the limit $B\to 0$.
Based on the preceding discussions, although the quasinormal frequencies are found to vary continuously, we concluded that the resultant tail would not converge to its massless counterpart.
This phenomenon is manifestly shown by Fig.~\ref{Fig9}.
It can be seen from the top plot of Fig.~\ref{Fig9} that as the effective mass decreases, the dissipative oscillations converge to the case of massless perturbations.
This is consistent with the numerical results obtained in Tab.~\ref{tab3}, where the Prony method~\cite{agr-qnm-16, agr-qnm-lq-matrix-06} has again been employed to extract the complex frequencies from the time profiles.
On the contrary, as shown in the bottom plot of Fig.~\ref{Fig9}, the late-time tail of massless perturbations is distinct from the remaining ones.
Specifically, the massless perturbation given by the orange curve does oscillate at a finite period, while the oscillation periods of massive perturbations are more significant (c.f., the green and dark red curves). 
Regarding the attenuation rate, all the curves of massive perturbations are essentially parallel, which is different from that of the massless one.
To be more quantitatively, the values extracted from the numerical fitting deviate from one another, as indicated in the caption of the plot.

We now argue that the case of massless perturbation corresponds to a scenario where the magnetic field was not there in the first place.
Such an interesting difference is somewhat reminiscent of the demagnetization process in ferromagnetic materials.
Mathematically, this can be understood as follows.
The convergence of the quasinormal frequencies can be understood as the analytic properties of the effective potential essentially remaining unchanged as the effective mass varies continuously.
However, the origin of a massive field's late-time tail is associated with the presence of two branch points on the real axis and the absence of any branch point at infinity.
As the effective mass approaches zero, the two branch points become degenerate, but there will not be a new branch cut to appear, which extends to infinity.
On the other hand, there is only one single branch point located at the origin in the case of a massless field.
The branch cut on the negative imaginary axis connects the origin with the second branch point at infinity.
As a result, the relevant structure on the frequency plane due to a massless field is intrinsically different from that of a massive perturbation, even though the mass is insignificant.
In other words, if one gradually removes the magnetic field from a magnetized black hole, the perturbation will evolve differently from the spacetime configuration where the magnetic field in question does not ever exist.
We note that the above picture is plausible only when the time scale of the demagnetization process is much smaller than the inverse of the typical distance between adjacent quasinormal frequencies.
This is because the resulting effective potential is now a function of time.
Therefore, the notion that QNMs correspond to the zeros of the Wronskian and subsequently the poles of the frequency domain Green's function~\cite{cqg-Nollert-qnm-charac} is only approximately valid when the process can be viewed as {\it quasistatic}. 
In literature, the master equation with time-dependent potential is mostly treated numerically~\cite{agr-qnm-time-dependent-03, agr-qnm-time-dependent-05, agr-qnm-lq-matrix-05}.

\section{Concluding remarks} \lb{section5}

To summarize, in the present study, we study the QNMs and late-time tails of charged massless scalar perturbations in a magnetized black hole in Rastall gravity.
It is shown that the magnetic field plays a significant role in both aspects of the resultant properties of the scalar perturbation.
To be specific, the massless scalar acquires an effective mass through the magnetic field, which significantly affects the temporal evolution of the initial perturbations.
For the quasinormal oscillations, the complex frequencies are distorted and might become either quasinormal resonance or unstable for specific parameters.
For the intermediate and late-time tails, power-law forms are obtained analytically and numerically, reminiscent of those of massive scalar perturbations.
Moreover, owing to the dynamic nature of the effective mass generated by the magnetic field, we argue that it possesses an interesting feature.
In particular, as one gradually reduces the external magnetic field, while the quasinormal frequencies converge to its massless counterpart, the behavior of the late-time tail deviates from the latter.
From a mathematical viewpoint, it can be understood by analyzing the structure of branch points and cuts on the complex frequency plane of the relevant Green's function.

\section*{Acknowledgments}
We wish to thank Rui-Hong Yue for enlightening discussions.
This work is supported by the National Natural Science Foundation of China (NNSFC) under contract Nos. 11805166, 11925503, and 12175076.
We also gratefully acknowledge the financial support from
Funda\c{c}\~ao de Amparo \`a Pesquisa do Estado de S\~ao Paulo (FAPESP),
Funda\c{c}\~ao de Amparo \`a Pesquisa do Estado do Rio de Janeiro (FAPERJ),
Conselho Nacional de Desenvolvimento Cient\'{\i}fico e Tecnol\'ogico (CNPq),
Coordena\c{c}\~ao de Aperfei\c{c}oamento de Pessoal de N\'ivel Superior (CAPES),
A part of this work was developed under the project Institutos Nacionais de Ci\^{e}ncias e Tecnologia - F\'isica Nuclear e Aplica\c{c}\~{o}es (INCT/FNA) Proc. No. 464898/2014-5.
This research is also supported by the Center for Scientific Computing (NCC/GridUNESP) of the S\~ao Paulo State University (UNESP).

\appendix
\section{Appendix}
\renewcommand{\theequation}{A.\arabic{equation}}
\setcounter{equation}{0}

In this Appendix, we show that when the magnetic field is small, the metric given by Eqs~\eqref{N3}-\eqref{N11} is a valid approximation.
For simplicity, one assumes the strength of the magnetic field is insignificant and does not consider the ``backreaction'' to the metric.
First, it is readily verified that Eq.~\eqref{N11} satisfies the Maxwell equation for linear electromagnetic field in Schwarzschild's solution of Einstein gravity.
For the Schwarzschild solution Eq.~\eqref{N10} of generalized Rastall gravity, one needs to ascertain that the deviation for Eq.~\eqref{N11} as linear electromagnetic field ($s=1$) is of higher-order, consistent with approximation taken later.
To show this, one rewrites the deviation from Eq.~\eqref{N21} as
\begin{equation}\label{N21prime}
\tilde f = f + \delta f ,
\end{equation}
and solves the Maxwell equation for $\delta f$.

It is not difficult to find that $\delta f$ satisfies
\begin{equation}\label{DeltafEq}
2\delta f + 2r\delta f' = \frac{6}{5r^3} ,
\end{equation}
which implies the magnitude of the correction $\delta f \sim O(r^{-3})$, irrelevant to the effective potential in the master equation Eq.~\eqref{N31}.

\bibliographystyle{h-physrev}
\bibliography{references_shao}

\end{document}